\catcode`\@=11
\newif\if@fewtab\@fewtabtrue
{\count255=\time\divide\count255 by 60
\xdef\hourmin{\number\count255}
\multiply\count255 by-60\advance\count255 by\time
\xdef\hourmin{\hourmin:\ifnum\count255<10 0\fi\the\count255}}
\def\ps@draft{\let\@mkboth\@gobbletwo
    \def\@oddhead{}
    \def\@oddfoot{\hbox to 7 cm{\tiny \versionno
       \hfil}\hskip -7cm\hfil\rm\thepage \hfil}
    \def\@evenhead{}\let\@evenfoot\@oddfoot}

\def\draftcite#1{\ifnum\draftcontrol=1#1\else{}\fi}

\def\@lbibitem[#1]#2{\item{}\hskip -3cm \hbox to 2cm
{\hfil$\scriptstyle\draftcite{#2}$}\hskip
1cm[\@biblabel{#1}]\if@filesw
     {\def\protect##1{\string ##1\space}\immediate
      \write\@auxout{\string\bibcite{#2}{#1}}}\fi\ignorespaces}

\def\@bibitem#1{\item\hskip -3cm \hbox to 2cm
{\hfil {\footnotesize\draftcite{#1}}}\hskip 1cm
\if@filesw \immediate\write\@auxout
       {\string\bibcite{#1}{\the\value{\@listctr}}}\fi\ignorespaces}

\def\citen#1{\if@filesw \immediate\write \@auxout {\string\citation{#1}}\fi%
\@tempcntb\m@ne \let\@h@ld\relax \def\@citea{}%
\@for \@citeb:=#1\do {\@ifundefined {b@\@citeb}%
    {\@h@ld\@citea\@tempcntb\m@ne{\bf ?}%
    \@warning {Citation `\@citeb ' on page \thepage \space undefined}}%
    {\@tempcnta\@tempcntb \advance\@tempcnta\@ne
    \setbox\z@\hbox\bgroup\ifcat0\csname b@\@citeb \endcsname \relax
    \egroup \@tempcntb\number\csname b@\@citeb \endcsname \relax
    \else \egroup \@tempcntb\m@ne \fi \ifnum\@tempcnta=\@tempcntb
    \ifx\@h@ld\relax \edef \@h@ld{\@citea\csname b@\@citeb\endcsname}%
    \else \edef\@h@ld{\hbox{--}\penalty\@highpenalty
    \csname b@\@citeb\endcsname}\fi
    \else \@h@ld\@citea\csname b@\@citeb \endcsname \let\@h@ld\relax \fi}%
\def\@citea{,\penalty\@highpenalty\hskip.13em plus.13em minus.13em}}\@h@ld}
\def\@citex[#1]#2{\@cite{\citen{#2}}{#1}}%
\def\@cite#1#2{\leavevmode\unskip\ifnum\lastpenalty=\z@\penalty\@highpenalty\fi%
  \ [{\multiply\@highpenalty 3 #1%
  \if@tempswa,\penalty\@highpenalty\ #2\fi}]}   %
\makeatother 

\def\ah            {{a_h^{}}}
\def\Ah            {\mbox{$^{\sssh}\!\!{\cal A}$}}
\def\ahp           {{a_{h'}^{}}}
\def\ainf          {\mbox{$_{}^{\sss(\infty)\!\!\!}{\cal A}$}}
\def\ainph         {\Halpha\iiN\ph}

\def\alg           {algebra}
\def\alh           {{a_{\el h}^{}}}
\def\alpham        {{\alpha_m}}
\def\ap            {{a_p^{}}}

\def\barw          {\overline w}
\def\barW          {\overline W}
\def\be            {\begin{equation}}
\def\bearl         {\begin{array}{l}}
\def\bearll        {\begin{array}{ll}}
\def\bepej         {\beta_{p_1}^{{\sss(}j{\sss)}}}
\def\bepij         {\beta_{p_i}^{{\sss(}j{\sss)}}}
\def\bepj          {\beta_p^{{\sss(}j{\sss)}}}
\def\bepn          {\beta_p^{{\sss(}n{\sss)}}}

\def\bah           {\mbox{$^{\sssh}\!{\cal B}$}}
\def\bak           {\mbox{$^{\sss(h')}\!{\cal B}$}}
\def\balh          {\mbox{$^{\ssslh}\!{\cal B}$}}
\def\bh            {{b_h^{}}}
\def\Bh            {{\tild b_h^{}}}
\def\Bhh           {{\tild b_{hh'}^{}}}
\def\bhp           {{b_{h'}^{}}}
\def\Bhp           {{\tild b_{h'}^{}}}
\def\bhz           {{b_h^{\sss(2)}}}
\def\binf          {\mbox{$^{\sss(\infty)\!\!}{\cal B}$}}
\def\Binf          {{}^{\sss(\infty)\!}{\cal B}}
\def\blh           {{b_{\el h}^{}}}

\def\bp            {{b_p^{}}}
\def\Bp            {{\tild b_p^{}}}
\def\bpn           {{b_{p^n}^{}}}
\def\Bpn           {{\tild b_{p^n}^{}}}
\def\bpne          {{b_{p^{n+1}}^{}}}
\def\Bpne          {{\tild b_{p^{n+1}}^{}}}
\def\bpe           {{b_{p_1^{}}^{}}}

\def\Bpene         {{\tild b_{p_1^{n_1}}^{}}}
\def\bpi           {{b_{p_i^{}}^{}}}
\def\Bpini         {{\tild b_{p_i^{n_i}}^{}}}
\def\bpp           {{b_{p'}^{}}}
\def\bpz           {{b_{p_2^{}}^{}}}

\def\calb          {\mbox{${\cal B}$}}
\def\cald          {{\cal D}}

\def\cat           {\mbox{${\cal C}$}}

\def\cft           {conformal field theory}
\def\cfts          {conformal field theories}

\def\Circ          {\!\circ\!}

\def\complex       {{\dl C}}
\def\Complex       {${\dl C}$}

\def\diago         {{\cal X}}
\newcommand{\D}[2] {\mbox{$D_{\!#1,#2}^{}$}}
\def\DD            {D}
\def\dl            {\mathbb }
\def\drac          {\displaystyle\frac }
\def\dsum          {\displaystyle\sum}

\def\eal           {\idem_\alpha}
\def\ealm          {\idem_\alpham}
\def\eals          {\langle\idem_\alpha\rangle}
\def\ee            {\end{equation}}

\def\eear          {\end{array}}
\newcommand\ef[1]  {\mbox{$f_{#1}^{}$}}
\newcommand\eF[1]  {{$\scs f_{#1}^{}$}}
\newcommand\efm[1] {\mbox{$f_{#1}^{-1}$}}
\newcommand\eg[1]  {\mbox{$g_{#1}^{}$}}
\newcommand\eG[1]  {{$\scs g_{#1}^{}$}}
\def\el            {\ell}
\def\eps           {\epsilon}
\def\epsh          {\epsilon_h}
\def\epsl          {\epsilon_{\el}}
\def\epsll         {\epsilon_{\el\el'}}
\newcommand\erf[1] {(\ref{#1})}
\def\ets           {\eta}
\newcommand\fb[1]  {\mbox{$\bar f_{#1}^{}$}}
\newcommand\ff[2]  {\mbox{$f^{}_{#1,#2}$}}
\newcommand\ffm[2] {\mbox{$f_{#1,#2}^{-1}$}}
\newcommand\fF[2]  {{$\scs f^{}_{#1,#2}$}}
\newcommand\FF[2]  {\scs f_{#1,#2}}
\def\fflh          {\ff{\el h}h}
\def\FFlh          {\FF{\el h}h}
\def\findim        {finite-dimensional}
\def\fline         {{~}\\[1 mm]\noindent ------------------\\[1 mm]}
\def\fns           {\footnotesize}
\def\frc           {fusion rule coefficient}
\def\fusg          {${\cal F}\!us\slb\g\srb$}
\def\fusgb         {${\cal F}\!u\!\!\overline{\!\!\mbox{~~}s\slb\g\srb}$}
\def\futnot#1      {\ifnum\draftcontrol=1%
                   \footnote{~{\sc internal footnote:} #1}\ \fi}
\def\futnote#1     {\footnote{~#1}\ }
\def\g             {\mbox{$\mathfrak g$}}
\def\gb            {\mbox{$\bar{\mathfrak g}$}}

\newcommand\ggt[2] {{\rm lcd}(#1,#2)}
 
\def\gp            {g}
\def\gv            {\mbox{$g_{}^{\scriptscriptstyle\vee}$}}
\def\halpha        {{\alpha(h)}}
\def\Halpha        {\alpha}
\def\halpham       {{\alpham(h)}}
\def\hbeta         {{\beta(h)}}
\def\Hbeta         {\beta}
\def\he            {{h_1}}
\def\hee           {\invmod\he\pene}
\def\hhpalpha      {{\alpha(hh')}}
\def\hi            {{h_i}}
\def\hip           {{h_{i}'}}
\def\hii           {\invmod\hi\pini}
\def\hiip          {\invmod\hip\pinip}

\def\ho            {{h_\circ}}
\def\hoe           {{h_{\circ,1}}}
\def\hoz           {{h_{\circ,2}}}
\def\hpalpha       {{\alpha(h')}}

\def\hrho          {\rho}
\def\hsa           {horizontal subalgebra}
\newcommand\hsp[1] {\mbox{\hspace{#1 mm}}}

\def\hwr           {highest weight representation}

\def\hy            {$\mbox{-\hspace{-.66 mm}-}$}
\def\hz            {{h_2}}
\def\hzz           {\invmod\hz\pznz}
\def\I             {\mbox{$\II$}}
\def\id            {{\sf id}}
\def\Id            {{\cal I}}
\def\idem          {{\rm e}}
\def\idemh         {{}^{\sssh\!}{\rm e}}
\def\ii            {{\rm i}}
\def\II            {I}
\def\iin           {\downarrow}
\def\iiN           {\!\downarrow\!\!}
\def\ihwm          {irreducible highest weight module}
\def\iN            {\!\in\!}
\def\indu          {{}^{\sss(\infty)\!\!}P}

\def\Infdim        {Infinite dimensional}
\def\ini           {\in\II}
\def\iNi           {\iN\II}
\def\irrep         {irreducible representation}
\newcommand\invmod[2] {\llb #1 \lrb^{-1}_{\displaystyle #2}}
\newcommand\jb[1]  {\mbox{$\bar\jmath_{#1}^{}$}}
\newcommand\jj[2]  {\mbox{$\jmath^{}_{#1,#2}$}}
\def\kma           {Kac\hy Moo\-dy algebra}
\def\kpf           {Kac\hy Peterson formula}
\def\kv            {\mbox{$k_{}^{\scriptscriptstyle\vee}$}}
\long\def\labl#1   {\label{#1}\ee \ifnum\draftcontrol=1 
                   \mbox{ }\\[-12 mm]\query{#1}\\[5 mm] \fi}
\def\l             {\preceq}
\def\lh            {\mbox{$^{\sssh\!\!}L$}}
\def\lhalpha       {{\alpha(\el h)}}

\def\Lhbeta        {\beta}
\def\lhs           {left hand side}
\def\lhrho         {\rho}
\def\lie           {Lie algebra}
\def\Lie           {Lie group}
\def\lk            {\mbox{$^{\sss(h')}\!L$}}
\def\llb           {\mbox{\large[}}
\def\lLb           {\mbox{\large(}}
\def\Llb           {\mbox{\Large\{}}
\def\LLb           {\mbox{\Large(}}
\def\llfont        {\sf }
\def\lrb           {\mbox{\large]}}
\def\lRb           {\mbox{\large)}}
\def\Lrb           {\mbox{\Large\}}}
\def\LRb           {\mbox{\Large)}}
\def\lrc           {Litt\-le\-wood\hy Ri\-chard\-son coefficient}
\def\lv            {\mbox{$\overline L^{\scriptscriptstyle\vee}$}}
\def\lV            {\bar L^{\scriptscriptstyle\vee}}
\def\lw            {\mbox{$\overline L^{\rm w}$}}
 
\def\Mid           {\!\mid\!}
\def\mm            {{\cal M}}
\def\mod           {\ {\rm mod}\;}

\newcommand\mult[1] {{\rm mult}^{}_{#1}}
\def\mw            {{\cal S}}
\def\mx            {\eta}
\def\mydollar      {{
}}
\def\mydollar      {$^\pounds$} 
\newcommand\n[3]   {\mbox{${\cal N}_{\!#1,#2}^{\ \ #3}$}}
\newcommand\N[1]   {\mbox{${\cal N}_{\!#1}$}}
\def\nabc          {\n abc}
\def\natnum        {{\dl N}}
\newcommand\nb[3]  {\mbox{$\overline{\cal N}_{\!#1,#2}^{\ \ #3}$}}
\def\nbabc         {\nb abc}
\newcommand\nh[3]  {\mbox{$_{}^{\sssh\!\!}{\cal N}_{\!#1,#2}^{\ \ #3}$}}
\def\nhabc         {\nh abc}
\newcommand\nlh[3] {\mbox{$_{}^{\ssslh\!\!}{\cal N}_{\!#1,#2}^{\ \ #3}$}}

\def\om            {\omega}
\def\Om            {\mbox{$\Omega$}}
\def\onedim        {one-dimensional}
\def\Onedim        {One-dimensional}

\renewcommand\P[2] {\mbox{$F_{\!{#1},{#2}}^{}$}}
\def\pb            {\mbox{$\bar P$}}
\def\ph            {\mbox{$_{}^{\sssh\!\!}P$}}
\def\pH            {_{}^{\sssh\!}P}
\def\phI           {\phi^{}}
\def\pho           {\mbox{$_{}^{\sss(\ho)\!\!}P$}}
\def\php           {\mbox{$_{}^{\ssshp\!\!}P$}}
\def\pene          {{p_1^{n_1}}}
\def\pini          {{p_i^{n_i}}}
\def\pjnj          {{p_j^{n_j}}}
\def\pinip         {{p_i^{n'_i}}}
\def\pjnjp         {{p_j^{n'_j}}}
\def\pznz          {{p_2^{n_2}}}
\def\pk            {\mbox{$_{}^{\sss(h')\!\!}P$}}
\def\pl            {\mbox{$\cal L$}}
\def\plh           {\mbox{$_{}^{\ssslh\!\!}P$}}
\def\plH           {_{}^{\ssslh\!}P}
\def\pllh          {\mbox{$_{}^{\sssllh\!\!}P$}}
\def\plp           {\mbox{$\cal O$}}
\def\pp            {\mbox{$_{}^{\sss(p)\!\!}P$}}
\def\PP            {F}
\def\prepro        {pre-image property}

\def\prodjh        {\prod_{\scriptstyle j\atop \scriptstyle p_j|h}}
\def\prodjhp       {\prod_{\scriptstyle j\atop \scriptstyle p_j|h'}}
\def\psia          {\psi_{}^{\sss(a)}}
\def\psib          {\overline\psi}
\def\psio          {\psi_\circ}
\def\psip          {\psi'}
\def\psipp         {\psi''}
\def\qdim          {quantum dimension}
\newcommand\qd[2]  {\mbox{${}^{\sss(#1)}\pi_{#2}$}}

\def\qfts          {quantum field theories}
\long\def\query#1{\hskip 0pt{\vadjust{\everypar={}\small\vtop to 0pt{\hbox{}%
     \vskip -13pt\rlap{\hbox to 50.0pc{\hfil{\vtop{\hsize=8pc\tolerance=6000%
     \hfuzz=.5pc\rightskip=0pt plus 3em\noindent#1}}}}\vss}}}}%
\newcommand\qz[1]  {\mbox{${\dl Q}(\zeta_{#1})$}}
\newcommand\qzm[1] {\mbox{${\dl Q}(\zeta_{M#1})$}}

\def\rb            {\mbox{$\overline{\ring}$}}
\def\rdot          {\star}

\def\rep           {representation}
\def\Rep           {Representation}
\def\resp          {respectively}

\def\ri            {\mbox{$^{\sss(i)\!}{\cal R}$}}
\def\rinf          {\mbox{$^{\sss(\infty)\!}{\cal R}$}}
\def\ring          {\mbox{${\cal R}$}}
\def\rj            {\mbox{$^{\sss(j)\!}{\cal R}$}}
\def\rh            {\mbox{$^{\sssh\!}{\cal R}$}}
\def\rhi           {\mbox{$^{\sss(h_i)\!}{\cal R}$}}
\def\rhO           {\rho}
\def\rHo           {\mbox{$^{\sss(\ho)\!}{\cal R}$}}
\def\rhp           {\mbox{$^{\sss(h')\!}{\cal R}$}}
\def\rhs           {right hand side}
\def\rk            {\mbox{$^{\sss(h')\!}{\cal R}$}}
\def\rkk           {\mbox{$^{\sss(k)\!}{\cal R}$}}

\def\rlh           {\mbox{$^{\ssslh\!}{\cal R}$}}
\def\sbinf         {\mbox{$\langle^{\sss(\infty)\!\!}{\cal B}\rangle$}}
\def\sbinfc        {\mbox{$\overline{\langle^{\sss(\infty)\!\!}{\cal B}
                   \rangle}$}}
\def\scs           {\scriptstyle}
\newcommand\secref[1] {section \ref{s.#1}}
\newcommand\sect[1] {\section{#1}\setcounter{equation}{0}}
\newcommand\Sect[2] {\sect{#1}\label{s.#2}
                   \ifnum\draftcontrol=1 \query{s.#2} \fi}
\def\semitimes     {\begin{picture}(9.4,8)\put(1.57,0.32){\line(0,1)
                   {4.4}}\put(0,0){$\times$} \end{picture} }
\def\sh            {\mbox{$^{\sssh}\!S$}}
\def\shhp          {\mbox{$^{\ssshhp}\!S$}}
\def\shp           {\mbox{$^{\ssshp}\!S$}}
\def\slb           {\mbox{\scriptsize(}}
\def\slh           {\mbox{$^{\ssslh}\!S$}}
\def\sign          {\mbox{sign}}
\newcommand\sm[2]  {\sM_{#1,#2}}
\def\sM            {\mbox{\rm s}}
\newcommand\Sm[2]  {\SM_{#1,#2}}
\def\SM            {\mbox{\rm S}}
\def\smat          {$S$-matrix}
\def\srb           {\mbox{\scriptsize)}}
\def\ssi           {semi-simple}
\def\sss           {\scriptscriptstyle}
\def\sssllh        {\scriptscriptstyle(\el\el'h)}
\def\ssslh         {\scriptscriptstyle(\el h)}
\def\sssh          {\scriptscriptstyle(h)}
\def\ssshhp        {\scriptscriptstyle(hh')}
\def\ssshp         {\scriptscriptstyle(h')}
\def\sumihh        {{\displaystyle\sum_{\scriptstyle i\atop
                   \scriptstyle p_i|h',\,\scriptstyle p_i\not{\:|}\,h}}}
\def\sumieh        {{\displaystyle\sum_{\scriptstyle i\ne1\atop 
                   \scriptstyle p_i|h}}}
\def\sumiezh       {{\displaystyle\sum_{\scriptstyle i\ne1,2\atop 
                   \scriptstyle p_i|h}}}
\newcommand\sumn[1] {{\displaystyle\sum_{#1=1}^{n}}}
\newcommand\sumne[1] {{\displaystyle\sum_{#1=1}^{n-1}}}
\newcommand\sumnee[1] {{\displaystyle\sum_{#1=1}^{n_1-1}}}
\newcommand\sumneee[1]{{\displaystyle\sum_{#1=n_1}^{n_1'-1}}}
\newcommand\sumnepe[1]{{\displaystyle\sum_{#1=1}^{n_1'-1}}}
\newcommand\sumnie[1] {{\displaystyle\sum_{#1=1}^{n_i-1}}}
\newcommand\sumnipe[1]{{\displaystyle\sum_{#1=1}^{n_i'-1}}}
 
\newcommand\sumph[1] {{\displaystyle\sum_{#1\in{}^{\sssh\!}P}}}
\newcommand\sumpH[1] {\sum_{#1\in^{\sssh\!}P}}
\newcommand\sumphp[1] {{\displaystyle\sum_{#1\in{}^{\ssshp\!}P}}}
 
\newcommand\sumplh[1] {{\displaystyle\sum_{#1\in{}^{\ssslh\!}P}}}

\def\th            {\mbox{$^{\sssh\!}T$}}
\def\threedim      {three-di\-men\-si\-o\-nal}
\def\tild          {\tilde}
\def\topo          {limit topology}  
\def\twodim        {two-di\-men\-si\-o\-nal}
\def\U             {{}^{\sss(\infty)\!}Q}

\def\UM            {{}^{\sss(\infty)\!}Q^-}
\newcommand\version[1] {\ifnum\draftcontrol=1 \typeout{}\typeout{#1}\typeout{}
                   \vskip3mm \centerline{\fbox{\tt DRAFT -- #1 -- \today}}
                   \vskip3mm \fi}

\def\vphi          {\varphi^{}}
\def\vphib         {\bar\varphi^{}}
\def\vphih         {{}^{\sssh\!}\varphi^{}}
\def\Vphih         {{}^{\sssh\!}\varphi}
\def\vphihp        {{}^{\ssshp\!}\varphi^{}}
\def\vphilh        {{}^{\ssslh\!}\varphi^{}}
\def\vphiho        {{}^{\sss(\ho)\!}\varphi^{}}
\def\wh            {\mbox{$^{\sssh\!}W$}}
\def\whp           {\mbox{$^{\sss(h')\!}W$}}
\def\wH            {{}^{\sssh}W}
\def\wlh           {\mbox{$^{\ssslh\!}W$}}

\def\wp            {\mbox{$^{\sss(p)\!}W$}}
\def\wrt           {with respect to }
\def\wrtt          {with respect to the }
\def\WZW           {Wess\hy Zumino\hy Witten}

\def\wzwt          {WZW theory}
\def\wzwts         {WZW theories}
\def\xih           {\hat\xi}
\def\zet           {{\dl Z}}
\def\Zet           {${\dl Z}$}

\def\zetplus       {\mbox{${\zet}_{>0}$}}

\def\bP            {\begin{picture}}
\def\bPo           {\begin{picture}(0,0)}
\def\eP            {\end{picture}}
\newcommand\pictriangleupddr[6] {\bP(100,90) 
        \put(46,70){#1} \put(-2,0){#2} \put(80,0){#3}
        \put(17,38){{#4}} \put(71,38){{#5}} \put(41,-4){{#6}} 
        \put(47,65){\vector(-1,-2){25}}
        \put(53,65){\vector(1,-2){25}}
        \put(25,5){\vector(1,0){50}}
        \eP}
\newcommand\pictriangledowndd[3] {\bP(100,90)
        \put(44,0){#1} \put(17,38){{\fns #2}} \put(72,38){{\fns #3}} 
        \put(47,15){\vector(-1,2){25}}
        \put(53,15){\vector(1,2){25}}
        \eP}

\global\def\draftcontrol{0}
\catcode`\@=12

\documentclass[12pt]{article} \usepackage{amssymb,amsfonts}

\begin{document}
\version\versionno
\setlength{\unitlength}{.1 em}

\begin{flushright}  
{\sf hep-th/9609124} \\ 
{\sf DESY 96-196} \\ {\sf IHES/P/96/47} \\
{\sf September 1996}
\end{flushright}
\begin{center} \vskip 1cm

{\Large\bf WZW FUSION RINGS IN THE LIMIT} \vskip 2mm
{\Large\bf  OF INFINITE LEVEL}  \vskip 12mm
{{\large \ J\"urgen Fuchs} \mydollar}\\[5mm] 
{\small DESY}\\ {\small Notkestra\ss e 85}\\ {\small D -- 22603~~Hamburg}
\\[5mm] and \\[5mm] {\large Christoph Schweigert}\\[5mm] {\small IHES}\\ 
{\small 35, Route de Chartres}\\ {\small F -- 91440 Bures-sur-Yvette}
\end{center}

\vskip 15mm

\begin{quote} {\bf Abstract.} \\
We show that the WZW fusion rings at finite levels form a projective system 
\wrtt partial ordering provided by divisibility of the height, i.e.\ the 
level shifted by a constant. {}From this projective system we obtain WZW 
fusion rings in the limit of infinite level. This projective limit constitutes 
a mathematically well-defined prescription for the `classical limit' of 
\wzwts\ which replaces the naive idea of `sending the level to infinity'.
The projective limit can be endowed with a natural topology, which plays
an important r\^ole for studying its structure. The representation theory
of the limit can be worked out by considering the associated fusion algebra;
this way we obtain in particular an analogue of the Verlinde formula.
\end{quote}
\vfill{}\fline{} {\small \mydollar~~Heisenberg fellow} \newpage

\Sect{Fusion rings}{wzw}

Fusion rings constitute a mathematical structure which emerges in various
contexts, for instance in the analysis of the superselection rules of \twodim\ 
quantum field theories; they describe in particular the basis independent 
contents of the operator product \alg\ of \twodim\ \cfts\ (for a review see 
\cite{jf24}). By definition, a {\em fusion ring\/} \ring\ is 
a unital commutative associative ring over the integers \Zet\ which possesses 
the following properties: there is a distinguished basis $\calb=\{\vphi_a\}$ 
which contains the unit and in which the structure constants \nabc\ are 
non-negative integers, and the evaluation at the unit 
induces an involutive automorphism, called the conjugation of \ring.
A fusion ring is referred to as {\em rational\/} iff it is \findim.
A rational fusion ring is called {\em modular\/} iff the matrix $S$ that 
diagonalizes simultaneously all fusion matrices \N a\ (i.e.\ the matrices 
with entries $(\N a)_b^{\ c} =\nabc$) 
is symmetric and together with an appropriate diagonal matrix $T$ generates a 
unitary \rep\ of SL$(2,\zet)$ (see e.g.\ \cite{kawA,jf24}).

In this paper we consider the fusion rings of (chiral, unitary) \wzwts. A 
\wzwt\ is a \cft\ whose chiral symmetry \alg\ is the semidirect sum of the
Virasoro \alg\ with an untwisted affine \kma\ \g, with the level \kv\ of the
latter a fixed non-negative integer. 
To any untwisted affine \kma\ \g\ we can thus associate a family of 
fusion rings, parametrized by the level \kv. The issue that we address in this
paper is to construct an analogue of the WZW fusion ring for infinite level,
which is achieved by giving a prescription for `sending the level to infinity' 
in an unambiguous manner. 

In view of the Lagrangian realization of \wzwts\ as sigma models, this 
procedure may be regarded as taking the `classical limit' of \wzwts.
Performing a classical limit of a parametrized family of \qfts\ 
is a rather common concept in the path integral formulation of quantum 
theories; it simply corresponds to sending Planck's constant to zero, and
hence provides a kind of correspondence principle. In the 
Lagrangian description of \wzwts\ as principal sigma models with
Wess\hy Zumino terms, the r\^ole of Planck's constant is played by the inverse
of the level \kv\ of the underlying affine \lie\ \g. However, it is known that 
the path integral of a WZW sigma model strictly makes sense only if the level 
\kv\ is an integer. In contrast to the path integral description, 
in the framework of algebraic approaches to quantum theory so far almost
no attempts have been made to investigate limits of \qfts. In this paper
we address this issue for the case of \wzwts. Now in an \alg ic treatment of 
\wzwts\ the integrality requirement just mentioned is immediately manifest. 
Namely, one observes that the structure of the theory
depends sensitively on the value \kv\ of the level. For non-negative integral
\kv\ the state space is a direct sum of unitary \ihwm s of the algebra \g, but 
its structure changes quite irregularly when going from \kv\ to $\kv+1$; at 
intermediate, non-integral, values of the level there do not even exist any
unitarizable \hwr s.

These observations indicate that it is a rather delicate issue to define what is 
meant by the classical limit of a \wzwt, and it seems mandatory to perform this 
limit in a manner in which the level \kv\ is manifestly kept integral (actually,
treating the level formally as a continuous 
variable is potentially ambiguous even in situations where one deals with
expressions which superficially make sense also at non-integral level).
It must also be noted that a priori it is by no means clear whether the so 
obtained limit will be identical with or at least closely resemble the 
structures which originally served to define the quantum theory in terms of
some classical field theory; in the case of WZW fusion rings, this
underlying classical structure is the \rep\ ring of the \findim\ simple \lie\
$\gb$ that is generated by the zero modes of the affine \lie\ \g. Indeed, it seems to be a quite generic feature of quantum theory that
the classical limit does not simply reproduce the classical structure one
started with. (Compare for instance the fact that in the path integral
formulation of quantum field theory the classical paths are typically of 
measure zero in the space of all paths that contribute to the path integral. 
Similar phenomena also show up when the continuum limit of a lattice theory 
is constructed as a projective limit; see e.g.\ \cite{asle2,bellst}.)
However, it is still reasonable to expect that the original
classical structure is, in a suitable manner, contained in the classical limit;
as we will see, this is indeed the case for our construction.

The desire of being able to perform a limit in which the level tends to
infinity stems in part from the fact
that \wzwts\ and their fusion rings can be used to define a regularization
of various systems (such as \twodim\ gauge theories or the Ponzano\hy Regge
theory of simplicial \threedim\ gravity), with the unregularized system 
corresponding, loosely speaking, to the classical theory. As removing the 
regulator is always a subtle issue, it is mandatory that the limit of 
the regularized theory is performed in a well-defined, controllable manner, 
which, in addition,
should preserve as much of the structure as possible.

\medskip
The basic idea which underlies our construction of the limit of WZW fusion rings
is to interpret the collection of WZW fusion rings as a category \fusg\ within
the category of all commutative rings and identify inside this category a 
projective system. By a standard category theoretic construction we can then 
obtain the limit (also known as the projective limit) of this projective system.
The partial ordering underlying the projective system is based on a divisibility
property of the parameter $\kv+\gv$ that together with the choice of the \hsa\
\gb\ characterizes the \wzwt\ (\gv\ denotes the dual Coxeter number of \gb;
the sum $\kv+\gv$ is called the {\em height\/}). 
In contrast, in the literature often a purely 
formal prescription `\,$\kv\to\infty$\,' is referred to as the classical
limit of \wzwts. In that terminology it is implicit that the standard ordering 
on the set of levels is chosen to give it the structure of a directed set. 
Now the projective limit is associated to a projective system as a whole,
not just to the collection of objects that appear in the system; in particular 
it depends on the underlying directed set and hence on the choice of partial 
ordering. Our considerations show, as a by-product, that it is not possible to
associate to the standard ordering any well-defined limit of the fusion rings.

The rest of this paper is organized as follows. We start in subsection
\ref{swz} by introducing the category \fusg\ of WZW fusion rings
associated to an untwisted affine \lie\ \g; 
in subsection \ref{s.qdim} conditions for the existence of 
non-trivial morphisms of this category are obtained.
In subsection \ref{s.ps} we define the projective system, and in the remainder
of section \ref{s.PS} we check that the morphisms introduced by this definition
possess the required properties. The projective limit of the so obtained 
projective system is a unital commutative associative ring of countably 
infinite dimension. This ring \rinf\ is constructed in \secref{pl}; there 
we also gather some basic properties of \rinf\ and introduce a natural
topology on \rinf. In \secref{binf} a concrete description of a 
distinguished basis \binf\ for the projective limit is obtained. This 
basis is similar to the distinguished bases
of the fusion rings at finite level; in order to show that \rinf\ is indeed
generated by \binf, the topology on \rinf\ plays an essential r\^ole. 
In \secref{gb} we demonstrate that the \rep\ ring of the \hsa\ 
$\gb\subset\g$ is contained in the projective limit \rinf\ as a proper subring.
In the final \secref{rep} we study the \rep\ theory of \rinf, \resp\ of the
associated fusion algebra over \Complex\,. In particular,
we determine all \irrep s and show that \rinf\ 
possesses a property which is the topological analogue of semi-simplicity,
namely that any continuous \rinf-module is the closure
of a direct sum of irreducible modules. To obtain these results it is again
crucial to treat the projective limit as a topological space.
Finally, we establish an analogue of the Verlinde formula for \rinf.

\Sect{The projective system of WZW fusion rings}{PS}

\subsection{WZW fusion rings}\label{swz}

The primary fields of a unitary \wzwt\
are labelled by integrable highest weights of the relevant affine \lie\
\g, or what is the same, by the value \kv\ of the level and by
dominant integral weights $\Lambda$ of \gb\ (the horizontal subalgebra of \g) 
whose inner product with the highest coroot of \gb\ is not larger than \kv.
We denote by \gv\ the dual Coxeter number of the simple \lie\ \gb\ and define
  \be  \I:= \{ i\in\zet \mid i\ge\gv \} \,.  \labl i
Thus \I\ is the set of possible values of the {\em height\/} $h\equiv \kv+\gv$
of the \wzwt\ based on \g. For any $h\iNi$ the fusion rules of a \wzwt\ at 
height $h$ define a modular fusion ring, with the elements of the distinguished 
basis corresponding to the primary fields. We denote this ring by \rh\ and its
distinguished basis by \bah, and the 
corresponding generators of SL$(2,\zet)$ by \sh\ and \th.

The distinguished basis \bah\ of the ring \rh\ can be labelled as 
  \be  \bah=\{\vphih_a \,|\,a\iN\ph\}  \ee
by the set 
  \be  \ph := \{ a\in\lw \mid a^i\ge1\ {\rm for}\ i=1,2,...\,,r;\;
  (a,\theta^{\scriptscriptstyle\vee})<h \} \labl{ph}
of integral weights in the interior of (the horizontal projection of) the
fundamental Weyl chamber of \g\ at {\em level\/} $h$; here $r$,
$\theta^{\scriptscriptstyle\vee}$ and \lw\ denote the rank, the highest coroot
and the weight lattice of \gb, respectively. Note that from
here on we use shifted \gb-weights $a=\Lambda+\rho$, which have level $h
=\kv+\gv$, in place of unshifted weights $\Lambda$ which are at level \kv. 
Here $\rho$ is the Weyl vector of \gb; in particular, $a=\rho$ is the label of 
the unit element
of \rh. This convention will simplify various formul\ae\ further on.

The ring product of \rh\ will be denoted by the symbol `\,$\rdot$\,'; thus
the fusion rules are written as
  \be  \vphih_a \rdot \vphih_b = \sumph c \nhabc\,\vphih_c \,. \ee

The collection $(\rh)_{h\ini}$ of such WZW fusion rings forms a 
category, more precisely a subcategory of the category of commutative rings,
which we denote by \fusg. The objects of \fusg\ are the rings \rh, 
and the morphisms (arrows) are those ring homomorphisms (which are automatically
unital and
compatible with the conjugation) which map the basis \bak\ up to sign factors 
to \bah. These are the natural requirements to be imposed on morphisms. Namely,
one preserves precisely all structural properties of the fusion ring,
except for the positivity of the structure constants; the latter
is not an algebraic property, so that one should be prepared to give it up.

\subsection{Existence of morphisms}\label{s.qdim}

It is not a priori clear whether the category \fusg\ as defined above has any 
non-trivial 
morphisms at all. To analyze this issue, we consider the quotients ${\sh_{a,b}}
/{\sh_{a,\rho}}$ ($a,b\iN\ph$) of \smat\ elements. These are known as the 
(generalized) \qdim s, or more precisely, as the $a$th \qdim\ of the element
$\vphih_b$, of the modular fusion ring \rh. The generalized \qdim s 
furnish precisely all inequivalent \irrep s of \rh\ \cite{kawA}. We denote by
  \be  \qd ha :\quad \rh\to\complex\,, \quad \vphih_b\mapsto 
  \frac{\sh_{a,b}}{\sh_{a,\rho}} \labl{qd}
the \irrep\ of \rh\ which associates to any element its $a$th generalized \qdim.

Assume now that $f\!:\;\rk\to\rh$ is a non-trivial morphism, i.e.\ a ring 
homomorphism which maps the distinguished basis \bak\ of \rk\ 
to the basis \bah\ of \rh. Then the composition $\qd ha\circ f$ provides us 
with a \onedim, and hence irreducible, \rep\ of \rk, i.e.\ we have
$\qd ha\circ f=\qd {h'}{a'}$ for some $a'\iN\pk$.
Let now \lh\ denote the extension of the field ${\dl Q}$ of rational numbers 
by the \qdim s\ $\sh_{a,b}/\sh_{a,\rho}$ of all elements of \bah; 
the observation just made then implies that
  \be  \lh \subseteq \lk  \labl{ll}
(when $f$ is surjective, one gets in fact the whole field $\lh$).
As we will see, this result puts severe constraints on the existence of 
morphisms from \rk\ to \rh. It follows from the \kpf\ \cite{KAc3} for the 
\smat\ that
  \be  \lh\subseteq\qzm h\,,  \labl q
with $\zeta_m:=\exp(2\pi\ii/m)$ and
$M$ the smallest positive integer for which all entries of the metric on the 
weight space of \gb\ are integral multiples of $1/M$
(except for $\gb=A_r$ where $M=r+1$, $M$ satisfies $M\leq4$). 
The inclusion \erf{ll} therefore implies that \lh\ lies in the intersection
$\qzm h \cap \qzm{h'} = \qzm{\,\ggt h{h'}}$,
and that this intersection is strictly larger than ${\dl Q}$ 
unless $\lh={\dl Q}$. Here $\ggt mn$ stands for the largest common divisor of 
$m$ and $n$. In the specific case that $h$ and $h'$ are coprime, $\ggt h{h'}=1$,
it follows that
  \be  \lh \,\subseteq\, \lk \cap \qzm h \,\subseteq\, \qzm{}  \,. \ee
Now typically the field \lh\ is quite a bit smaller than \qzm h,
i.e.\ the inequality \erf q is not saturated (e.g.\ if the ring is
self-conjugate, \lh\ is already contained in the maximal real subfield of \qzm 
h); nevertheless,
inspection shows that the requirement $\lh\subseteq\qzm{}$ is fulfilled only
in very few cases (for instance, for \gb\ of type $B_{2n},\,C_r,\,D_{2n},\,
E_7,\,E_8$ or $F_4$, one has $M\le2$ so that the requirement is just $\lh=
{\dl Q}$). In addition, the main \qdim s $\sh_{a,\rho}/\sh_{\rho,\rho}$ lie 
in fact in \qz{2h}, and hence the above requirement would restrict them to
lie in $\qz{2h}\cap\qzm{}=\qz{\ggt{2h}M}$, and thus to be rational whenever
$2h$ and $M$ are coprime.

It follows that for almost all pairs $h,\,h'$ of coprime heights there
cannot exist any morphism from \rk\ to \rh. The same arguments also
show that the existence of non-trivial morphisms becomes the more probable 
the larger the value of $\ggt h{h'}$ is. The most favourable situation is when
$h'$ is a multiple of $h$; in the next section we will show that in this
case a whole family of morphisms from \rk\ to \rh\ (with $h\iNi$ arbitrary)
can be constructed in a natural way.

The considerations above indicate in particular that the naive way of
taking the limit `\,$k\to\infty$\,' with the standard
ordering on the set \I\ cannot correspond to any well-defined limit of the
WZW fusion rings. In contrast, as we will show, when replacing the standard 
ordering by a suitable partial ordering, a limit can indeed be constructed,
namely as the projective limit of a projective system that is associated to
that partial ordering.

Let us also mention that the required ring homomorphism property implies 
that any morphism of \fusg\ maps simple currents to simple currents. 
(By definition, simple currents are those elements
$\varphi_a$ of the distinguished basis \bah\ which have inverses in the 
fusion ring; they satisfy $\sum_c\nabc=1$ for all $b$. Such elements are
sometimes also called units of the ring, not to be confused with the unit
element of the fusion ring.)

\subsection{The projective system}\label{s.ps}

On the set \I\ \erf i of heights one can define a partial ordering `\,$\l$\,' by
  \be  i \l j  \ :\Leftrightarrow\  i\,|\,j \,,  \labl l
where the vertical bar stands for divisibility.
For any two elements $i,i'\iNi$ there then exists a $j\iNi$ (for example, the
smallest common multiple of $i$ and $i'$) such that $i\l j$ and $i'\l j$.
Thus the partial ordering \erf l endows \I\ with the structure of a
{\em directed set}.

We will now show that
when the set \I\ is considered as a directed set via the partial ordering 
\erf l, the collection $(\rh)_{h\ini}$ of WZW fusion rings can be made into a
{\em projective system}, that is,
for each pair $i,j\iNi$ satisfying $i\l j$ there exists a morphism
  \be  \ff ji:\quad \rj\to\ri\,,  \labl f
such that \ff ii is the identity for all $i\iN I$ and
such that for all $i,j,k\iNi$ which satisfy $i\l j\l k$, the diagram
  \be  \bearl {}\\[-2.2em]
  \pictriangleupddr {$\!\!\!$\rkk}{\rj~~}{\ri}{\fF kj}{\fF ki}{\fF ji}
  \\[.3em] \eear \labl1
commutes.

We have to construct the maps \ff ij for all pairs $i,\,j$ with $i|j$. Writing 
$i=h$ and $j=\el h$ with $\el\iN\natnum$, the construction goes as follows.
The horizontal projection \wh\ of the affine Weyl group at height $h$ has 
the structure of a semidirect product $\wh = \barW\semitimes h\lv$,
with $\barW$ the Weyl group and \lv\ the coroot lattice of \gb, so that 
in particular \wh\ is contained
as a finite index subgroup in \wlh, the index having the value $\el^{\,r}_{}$.
Thus any orbit of \wh\ decomposes into orbits of \wlh, and each Weyl chamber 
at height $\el h$ is the union of $\el^{\,r}_{}$ Weyl chambers at height $h$.
As a consequence, we find that the following statement holds for the set \plh\ 
defined according to \erf{ph}.
To any $a\iN\plh$ there either exists a unique element $w_a\iN\wh$ such that
  \be  a':=w_a(a)  \labl{a'}
belongs to the set \ph, or else $a$ lies on the boundary of some affine Weyl 
chamber at height $h$. In the former case we define
  \be  \fflh(\vphilh_a):= \epsl(a)\cdot \vphih_{a'}   \labl'
with $\epsl(a)=\sign(w_a),$ while in the latter case we set
$\fflh(\vphilh_a):=0$. It is convenient to include this latter case  
into the formula \erf', which is achieved by setting
  \be  \epsl(a):= \left\{ \begin{array}{ll}
  0 & \mbox{if $a$ lies on the boundary of an}\\
           & \mbox{affine Weyl chamber at height $h$\,,}
  \\[1mm] \sign(w_a) & {\rm else} \,. 
  \end{array}\right. \labl{eps}

\subsection{Proof of the morphism properties}

We have to prove that $\ff ij$ defined this way is a 
ring homomorphism and that it satisfies the composition property \erf1. 
It is obvious from the definition that $\ff ii=\id$ (and also that
$\ff ij$ is surjective). To show the homomorphism property, we write $\ff ij$ 
in matrix notation, and for convenience use 
capital letters for the fusion ring \rlh\ and lower case letters for the
fusion ring \rh. Thus the elements of the basis \balh\ of \rlh\ are denoted 
by $\phI_A\equiv\vphilh_{\!A}$ with $A\iN\plh$, while for the elements
of \bah\ we just write $\vphi_a$ with $a\iN\ph$, and we use the notation \SM\
and \sM\ for the $S$-matrices in place of \slh\ and \sh, respectively. 
The mapping is then defined on the preferred basis $\balh$ as 
  \be  \fflh(\phI_A) = \sumph b \P Ab\, \vphi_b  \,  \labl3
with
  \be  \P Ab \equiv {}^{\sss(\el h,h)\!\!}_{}\P Ab := \epsl(A)\,
  \delta^{}_{w_A(A),b}  \,,  \labl P
and extended linearly to all of \rlh. As has been established in 
\cite{fusS2}, the matrix \erf P satisfies the relations
 \futnote{In \cite{fusS2}, mappings of the type \erf' were encountered as
so-called quasi-Galois scalings. In that setting, the level of the \wzwt\ is
not changed, while the weights $A$ are scaled by a factor of $\el$, followed by
an appropriate affine Weyl transformation to bring the weight $\el A$ back to
the Weyl chamber \plh\ or to its boundary. Since what matters is only the 
relative
`size' of weights and the translation part of the Weyl group, these mappings
are effectively the same as in the present setting where there is no scaling
of the weights but the extension from \wlh\ to \wh\ scales the translation
lattice down by a factor of $\el$.
\\[.2em]Note that in \cite{fusS2} the letter $P$ was used for the matrix \erf P
in place of $\PP$, and $\DD$ was defined as the transpose of the matrix \erf D.}
  \be  \SM\,\PP = \el^{\,r/2}_{}\,\DD\,\sM \,, \qquad
  \PP\,\sM = \el^{\,r/2}_{}\,\SM\,\DD  \,, \labl{sS}
with
  \be  \D Ab \equiv {}^{\sss(\el h,h)\!\!}_{}\D Ab:= \delta_{A,\el b}\,. \labl D
Furthermore, from the Kac\hy Peterson formula \cite{KAc3} for the modular
matrix $S$, one deduces the identity
  \be  \sm ab = \el^{\,r/2}_{}\, \Sm{\el a}b  \labl K
for all $a,b\iN\ph$. 

Combining the relations \erf P -- \erf K and the 
Verlinde formula \cite{verl2}, we obtain for any pair $A,B\iN\plh$
  \be  \begin{array}{l}  \!\!
  \fflh(\phI_A \rdot \phI_B) = \sumplh C \nlh ABC\, \fflh(\phI_C)
  = \sumplh{C,D}\;\sumph e \drac {\Sm AD^{} \Sm BD^{} \Sm CD^*}{\Sm\rhO D}
    \, \P Ce\, \vphi_e
  \\{}\\[-2.6mm] \hsp{23.2}
  = \!\sumplh{D}\;\sumph e \drac {\Sm AD \Sm BD (\SM^*\!\PP)_{D,e}}{\Sm\rhO D} 
    \, \vphi_e
  = \el^{\,r/2}_{}\cdot \!\!\!\sumplh{D}\;\sumph e \drac {\Sm AD^{} \Sm BD^{} 
    (\DD\sM^*)_{D,e}}{\Sm\rhO D} \, \vphi_e
  \\{}\\[-2.6mm] \hsp{23.2}
  = \el^{\,r/2}_{}\cdot \sumph{d,e} \drac {\Sm A{\el d}^{} \Sm B{\el d}^{}
    \,\sm de^*} {\Sm \rhO{\el d}}\, \vphi_e
  = \el^{\,r}_{}\cdot \sumph{d,e} \drac {\Sm A{\el d}^{} \Sm B{\el d}^{}
    \,\sm de^*} {\sm \rhO d}\, \vphi_e
  \\{}\\[-2.6mm] \hsp{23.2}
  = \el^{\,r}_{}\cdot \sumph{d,e} \drac {(\SM\DD)_{A,d}^{} (\SM\DD)_{B,d}^{} \,
    \sm de^*} {\sm \rhO d}\, \vphi_e
  = \sumph{d,e} \drac {(\PP\sM)_{A,d}^{} (\PP\sM)_{B,d}^{} \,\sm de^*} 
    {\sm \rhO d}\, \vphi_e
  \\{}\\[-2.6mm] \hsp{23.2}
  = \sumph{a,b,d,e} \P Aa \P Bb\, \drac {\sm ad^{} \,\sm bd^{} \,\sm ed^*}
    {\sm \rhO d}\, \vphi_e
  = \sumph{a,b,c} \P Aa \P Bb\, \nhabc \, \vphi_c
  \\{}\\[-2.2mm] \hsp{23.2}
  = \sumph{a,b} \P Aa \P Bb\, \vphi_a \rdot \vphi_b
  = \fflh(\phi_A) \rdot \fflh(\phi_B)
  \end{array} \ee
Thus $\fflh$ is indeed a homomorphism.

As a side remark, let us mention that an analogous situation arises for the 
\cfts\ which describe a free boson compactified on a circle of rational radius
squared. These theories are labelled by
an (even) positive integer $h$, and for each value of $h$ the
fusion ring is just the group ring $\zet\zet_h$ of the abelian group
$\zet_h=\zet/h\zet$. The modular \smat\ is given by
$\sh_{p,q} = h^{-1/2} \exp(2\pi\ii p q /h)$, where the labels $p$ and $q$ 
which correspond to the primary fields are integers
which are conveniently considered as defined modulo $h$. It is straightforward 
to check that the identities \erf{sS} are again valid (with $r$ set to 1)
if one defines ${}^{\sss(\el h,h)\!}_{}\P Ab:=\delta^{(h)}_{A,b}$ and 
${}^{\sss(\el h,h)\!}_{}\D Ab:=\delta^{(\el h)}_{A,\el b}$, where the
superscript on the $\delta$-symbol $\delta^{(m)}_{a,b}$ indicates that 
equality needs to hold only modulo $m$. As a consequence, this way we obtain
again a projective system based on the divisibility of $h$ (the composition
property is immediate). Moreover, precisely as in the case of \wzwts, 
with a different partial ordering of
the set $\{h\}=\zetplus$ it is not possible to define a projective system.

\subsection{Proof of the composition property}

Finally, the composition property \erf1 of the homomorphisms \erf' is
equivalent to the relation
  \be  \sumplh B
  {}^{\sss(\el\el'h,\el h)\!}_{}\P {\llfont A}B \,
  {}^{\sss(\el h,h)\!}_{}\P Bc =
  {}^{\sss(\el\el'h,h)\!}_{}\P {\llfont A}c \labl{ppp}
among the projection matrices $\PP$ that involve the three different
heights $h$, $\el h$ and $\el\el'h$.
Here as before the elements of \ph\ and \plh\ are denoted by lower
case and capital letters, respectively, while for the elements of \pllh\
we use sans-serif font. The relation \erf{ppp} is in fact an immediate 
consequence of the definition of
the homomorphisms \ff ij. The explicit proof is not very illuminating;
the reader who wishes to skip it should proceed directly to \secref{pl}.

To prove \erf{ppp}, let us first assume that the left hand side does not
vanish. Then there exist unique Weyl transformations $\barw_1,\barw_2
\iN\barW$ and unique vectors $\beta_1,\beta_2\iN h\lv$ in the 
coroot lattice scaled by $h$, and a unique weight $B\iN\plh$, such that
  \be  \barw_1({\llfont A}) + \el\beta_1 =  B \,, \qquad
  \barw_2(B) +\beta_2 = c \,,   \labl{ww}
and the left hand side of \erf{ppp} takes the value
  \be  \epsll({\llfont A})\,\epsl(B) = \sign(\barw_1)\,\sign(\barw_2) 
  = \sign(\barw_1\barw_2) \,. \labl{sww}
By combining the two relations \erf{ww}, it follows that
  \be  \barw_2\barw_1({\llfont A}) + \beta = c \,, \labl{beta}
where $\beta=\el\,\barw_2(\beta_1)+\beta_2$. Since $\beta$ is again an  
element of $h\lv$, this means that \erf{beta} describes, up to sign, the
mapping corresponding to the right hand side of \erf{ppp}. Further, the
sign of the right hand side is then given by $\sign(\barw_2\barw_1)$
and hence equal to \erf{sww}; thus \erf{ppp} indeed holds.

We still have to analyze \erf{ppp} when its left hand side vanishes.
Then either the $\llfont A$\,th row of ${}^{\sss(\el\el'h,\el h)\!}_{}\PP$ 
or the $c$\,th column of ${}^{\sss(\el h,h)\!}_{}\PP$ must be zero.
In the former case, the weight ${\llfont A}\iN\pllh$ belongs to the
boundary of some Weyl chamber \wrt \wlh, and thus there exist $\barw\iN\barW$
and $\beta\iN h\lv$ such that $\barw({\llfont A})+\el\beta={\llfont A}$. 
But this means that ${\llfont A}\iN\pllh$ also lies on the
boundary of some Weyl chamber \wrt $\wh\supset\wlh$, and hence also
the right hand side of \erf{ppp} vanishes as required.
In the second case, there are unique elements $w_1\iN\wlh$ and $w_2\iN\wh$
satisfying $w_1({\llfont A})=B$ and $w_2(B)=B$.
Because of $\wlh\subset\wh$, $w_1$ 
can also be considered as an element of the Weyl group \wh\ at height $h$. 
By assumption, $w_2$ is a non-trivial element of \wh. The product
$w_0:=w_1^{-1}w_2^{}w_1^{}\iN\wh$ is then non-trivial, too, and satisfies
  \be  w_0({\llfont A}) =w_1^{-1}w_2^{}w_1^{}({\llfont A})
  =w_1^{-1}w_2^{}(B)=w_1^{-1}(B)={\llfont A} \,. \ee
Thus the weight $\llfont A$ is invariant under a non-trivial element of \wh\
and hence lies on the boundary of some Weyl chamber \wrt \wh; this implies
again that the right hand side of \erf{ppp} vanishes as required.

This concludes the proof of \erf{ppp}, and hence of the claim that
$(\rh)_{h\ini}$ together with the maps \ff ij constitutes a projective system.

\Sect{The projective limit \rinf}{pl}

We are now in  a position to construct the projective limit \rinf\ of
the projective system that we introduced in subsection \ref{s.ps}. 

\subsection{Projective limits and coherent sequences}

A projective system $(\rh)_{h\in I}$\, in some category \cat\ is said to
possess a {\em projective limit\/} $(\pl,f)$ (also called 
the inverse limit, or simply the limit) if there exist an object \pl\
as well as a family $f$ of morphisms $\ef h\!:\ \pl\to\rh$ (for all $h\iN\I$)
which satisfy the following requirements (see e.g.\ \cite{ENCY}).
First, for all $h,h'\iN\I$ with $h\l h'$ the diagram
  \be  \bearl {}\\[-2.2em]
  \pictriangleupddr {\pl}{\rk~~}{\rh}{\eF {h'}}{\eF h}{\fF{h'}h} 
  \\[.3em] \eear \labl I
commutes; and second, the following {\em universal property\/} holds:
for any object \plp\ of the category for which a family of morphisms 
$\,\eg h\!:\;\plp\to\rh\,$ ($h\iN\I$) exists which possesses
a property analogous to \erf I, i.e.\
  \be  \ff {h'}h \circ \eg{h'} = \eg h  \quad {\rm for}\ \, h\l h' \, , \ee
there exists a unique morphism $\,\gp\!:\;\plp\to\pl$ such that the diagram
  \be \bearl \bP(120,170)
  \put(0,90){ \pictriangleupddr {\pl}{\rk~~}{\rh}{\eF{h'}}{\eF h}{\fF{h'}h} }
  \put(0,18){ \pictriangledowndd {\,\plp}{\eG{h'}}{\eG h} }
  \put(113,92){\bPo \put(0,0){\oval(30,142)[r]} \put(19.8,3.5){$\scs\gp$}
      \put(0,71){\vector(-1,0){45}} \put(0,-71){\line(-1,0){45}} \eP}
  \eP \\[-1.8em] \eear \labl{II}
commutes for all $h,h'\iN\I$ with $h\l h'$.

To be precise, in the above characterization of the projective limit $(\pl,f)$
it is implicitly assumed that \pl\ is an object in \cat\ and that the \ef i are
morphisms of \cat. But in fact such an object and such morphisms need not exist.
In the general case one must rather employ a definition of the projective
limit as a certain functor from the category \cat\ to the category of sets,
and then the question arises whether this functor is `representable' through
an object \pl\ and morphisms \ef i as described above. In this language
the crucial issue is the existence of a representing object \pl\
(see e.g.\ \cite{ARti,PAre,HIst}). 
\futnot{PAre': sec. 2.5; HIst: chap. VIII.5}

Now one and the same projective system can frequently 
be regarded as part of various different categories. 
For instance, when describing the projective system of our interest one can
restrict oneself to the category \fusg. As we will see, when doing so
a projective limit of the projective system does not exist. But
one can also consider it, say, in the category of commutative rings, or in the 
still bigger category of vector spaces, or even in the category of
sets. The existence and the precise form of the projective limit usually
depend on the choice of category.
In our case, however, the category \cat \,=\,\fusg\ we start with
is small, i.e.\ its objects are sets, and as a consequence
there exists a natural construction by which the object \pl\ and 
the morphisms \ef i can be obtained in a concrete manner (in particular, \pl\
is again a set). Moreover, it turns out that the projective limit we obtain in 
the category of sets is exactly the same as the limit
that we obtain in the category of commutative rings or vector spaces, 
which also indicates that this way of performing the limit is a quite natural.

This construction proceeds as follows.
Given a projective system of objects \rh\ and morphisms \ff{h'}h of a
small category \cat,
one regards the objects $\rh\iN\cat$ as sets and considers the infinite direct
product $\prod_{h\in\II}\!\rh$ of all objects of \cat.
The elements of this set are those maps 
  \be  \psi:\ \; \I \;\to\; \bigcup_{h\in\II}^. \rh   \labl{psi}
from the index set \I\ to the disjoint union of all objects \rh\
which obey $\psi(h)\iN\rh$ for all $h\iN\I$; 
they are sometimes called `generalized sequences' (ordinary sequences can be
formulated in this language by considering the index set ${\dl N}$ with the 
standard ordering $\le$\,). The subset $\rinf \subset\prod_{h\in\II}\!\rh$
consisting of {\em coherent sequences}, i.e.\ of those generalized sequences 
for which 
  \be  \ff {h'}h \circ \psi(h') = \psi(h)  \labl j
for all $h,h'\iN\I$ with $h\l h'$, is isomorphic to the projective limit.
More precisely, \rinf\ is isomorphic to \pl\ as a set, and the morphisms 
$\ef h$ are the projections to the components, i.e.\ 
  \be  \ef h(\psi):=\psi(h) \,. \ee

For the projective system introduced in subsection \ref{s.ps} where \cat\ is
the (small) category \fusg, the projective limit is clearly {\em not\/} 
contained in the original category, because no object \ri\ of \fusg\ 
can possess morphisms to {\em all\/} objects \rj. In order to identify
nevertheless a projective limit associated to the projective system
defined by \erf', it is therefore necessary to consider 
the set \rinf\ of coherent sequences. In accordance with the remarks above,
for definiteness from now on we will simply refer to \rinf\ as `the'
projective limit of the system \erf1 of WZW fusion rings.

\subsection{Properties of \rinf}

Let us list a few simple properties of the projective limit \rinf. First, \rinf\ 
is a ring over \Zet.  The product $\psi_1\rdot \psi_2$ in \rinf\ is defined 
pointwise, i.e.\ by the requirement that
  \be  (\psi_1\rdot \psi_2)(h):= \psi_1(h)\rdot \psi_2(h)  \labl{pcp}
for all $h\iNi$. This definition makes sense, i.e.\
for all $\psi_1,\psi_2\iN\rinf$ also their product is in \rinf, because
  \be  \begin{array}{ll}
  \ff{h'}h\circ(\psi_1\rdot \psi_2)(h') \!\!&=
  (\ff{h'}h\circ\psi_1(h'))\rdot  (\ff{h'}h\circ\psi_2(h')) \\[1.9mm]&
  = \psi_1(h) \rdot  \psi_2(h) = (\psi_1\rdot \psi_2)(h) \,; \end{array}\ee
here in the first line the morphism property of the maps $\ff{h'}h$ is used.
{}From the definition \erf{pcp} it is clear that the product of \rinf\ is 
commutative and associative, and that \rinf\ is unital, with the unit element 
being the element $\psio\iN\rinf$ that satisfies
  \be  \psio(h)=\vphih_{\!\rho}  \ee
for all $h\iNi$. 

Second, a conjugation $\psi\mapsto \psi^+$ can be defined on \rinf\ by setting
  \be  \psi^+(h):=(\psi(h))^+  \ee
for all $h\iNi$. The conjugation $\vphih\mapsto(\Vphih)^+$ on the rings \rh\ 
commutes with the projections $\ff{h'}h$. As a consequence, indeed 
$\psi^+\iN\rinf$ whenever $\psi\iN\rinf$, conjugation is an involutive
au\-to\-morphism of \rinf, and the unit element $\psio$ is self-conjugate.

In \secref{binf} we will construct a countable basis \binf\ of the ring \rinf;
this basis contains in particular the unit element $\psio$.
For any $\psi\iN\binf$ and any $h\iNi$, $\psi(h)$ is either zero or, up to 
possibly a sign, an element of the basis \bah\ of \rh.
Also, while by construction the structure constants in the basis \binf\
are integers, there seems to be no reason why they should be non-negative.
Accordingly, an interpretation of the limit \rinf\
as the \rep\ ring of some underlying \alg ic structure is even less obvious
than in the case of the fusion rings \rh.\,%
\futnote{The latter can e.g.\ be regarded as the \rep\ rings of the `quantum 
symmetry' of the associated \wzwts. However, so far there is no agreement on 
the precise nature of those quantum symmetries.}
In particular, in \secref{gb} we will see that \rinf\ does not coincide
with the \rep\ ring \rb\ of the simple \lie\ $\gb\subset\g$, but rather
that it contains \rb\ as a tiny proper subring.

As it turns out, the fusion product of two elements of \binf\ is generically
{\em not\/} a finite linear combination of elements of \binf, or in other
words, \binf\ does not constitute an ordinary basis of \rinf. Rather, it must
be regarded as a topological basis. For this interpretation to make sense,
a suitable topology on \rinf\ must be defined. This will be achieved in
the next subsection.

\subsection{The \topo\ of \rinf}\label{topo}

The fusion rings \rh\ can be considered as topological spaces by simply
endowing them with the discrete topology, i.e.\ by declaring every subset to be
open (and hence also every subset to be closed). The projective limit \rinf\ 
then becomes a topological space in a natural manner, namely by defining its 
topology as the coarsest topology in which all projections $\ef h\!:\;
\rinf\to\rh$ are continuous; this will be called the {\em\topo\/} on \rinf.
\futnot{This is the natural adaptation of the product topology on the
direct product $\prod_{h\in\II}\rh$.}

More explicitly, the \topo\ on \rinf\ is described by the property that
any open set in \rinf\ is an (arbitrary, i.e.\ not necessarily finite nor
even countable) union of elements of
  \be  \Om:= \{ \efm h(M) \mid h\iN\I,\; M\subseteq\rh \} \,, \labl{Om}
i.e.\ of the set of all pre-images of all sets in any of the fusion rings \rh.\\
Note that we need not require to take also finite intersections of these
pre-images. This is because \Om\ is closed under taking finite intersections,
as can be seen as follows. Let $\om_i\iN\Om$ for $i=1,2,...\,,N$; by definition,
each of the $\om_i$ can be written as $\om_i=\efm{h_i}(M_i)$ for some heights
$h_i\iNi$ and some subsets $M_i\subseteq\rhi$. Denote then by $h$ the smallest
common multiple of the $h_i$ for $i=1,2,...\,,N$. Because of \erf I we have
$\ef{h_i}=\ff h{h_i}\Circ\ef h$, so that
  \be  \efm{h_i}(M_i)=\efm h(\ffm h{h_i}(M_i))=\efm h(\tilde M_i) \,, \ee
where for all $i=1,2,...\,,N$ the sets $\tilde M_i:=\ffm h{h_i}(M_i)$ 
are subsets of \rh.
Because of $\bigcap_{i=1}^N\tilde M_i\subseteq \rh$, it thus
follows that 
  \be  \bigcap_{i=1}^N \om_i =\bigcap_{i=1}^N \efm h(\tilde M_i)
  = \efm h\llb\bigcap_{i=1}^N \tilde M_i\lrb  \ee
is an element of the set \Om\ \erf{Om}. Thus \Om\ is closed under taking 
finite intersections, as claimed.\\
As a consequence of this property of \Om, in particular any non-empty open 
set in \rinf\ contains a subset which is of the form $\efm h(M)$ for some 
$h\iNi$ and some $M\subseteq\rh$; for later reference, we call this fact
the `\prepro' of the non-empty open sets in \rinf.

Note that the \topo\ on \rinf\ is {\em not\/} the discrete one, but finer. 
To see this, suppose the \topo\ were the discrete one. Then for any $\psi\iN
\rinf$ the one-element set $\{\psi\}$ would be open and hence a
union of sets in $\Omega$ \erf{Om}; but as $\{\psi\}$ just contains one single 
element, this means that it even has to belong itself to $\Omega$.
This in turn means that there would exist $h\iNi$ and $M\subseteq\rh$ such that
$\{\psi\}=f_h^{-1}(M)$, and hence simply $M=\{\psi(h)\}$. This, however, would
imply that each element $\psi\iN\rinf$ would already be determined uniquely by
the value $\psi(h)$ for a single height $h$. {}From the explicit description 
of \rinf\ as a space of coherent generalized sequences,
it follows that this is definitely not true. Thus the 
assumption that the \topo\ is the discrete one leads to a contradiction.

Whenever two elements $\psi,\psi'\iN\rinf$ are distinct, there exists some 
height $h\iNi$ such that $\psi(h)\ne\psi'(h)$. The open subsets
$\omega:=f_h^{-1}(\{\psi(h)\})$ and $\omega':=f_h^{-1}(\{\psi'(h)\})$ then
satisfy $\psi\iN\omega$ and $\psi'\iN\omega'$ as well as
$\omega\cap\omega'=\emptyset$. This means that when endowed with the \topo,
\rinf\ is a Hausdorff space.

\Sect{A distinguished basis \binf\ of \rinf}{binf}

In this section we construct a (topological)
basis \binf\ of the projective limit \rinf\ of WZW fusion rings. 

\subsection{A linearly independent subset of \rinf}\label{s1}

We start by defining the subset $\binf\subset\rinf$ as the set of all those 
elements $\psi\iN\rinf$ which for every $h\iNi$ satisfy
  \be  \psi(h)= \epsh\cdot\vphih_a  \labl{bd}
for some 
  \be  a\iN\ph \qquad{\rm and}\qquad \epsh\iN\{0,\pm1\}  \ee
(i.e.\ for each height
$h$ the fusion ring element $\psi(h)\iN\rh$ is either zero or, up to a sign, an
element of the distinguished basis \bah), and for which in addition not all of 
the prefactors $\epsh$ vanish and $\epsh=1$ for the smallest $h\iNi$ for which
$\epsh\ne0$. The latter requirement ensures that $-\psi\not\in\binf$ for
all $\psi\iN\binf$.

Note that at this point we cannot tell yet whether the set \binf\ is large
enough to generate the whole ring \binf; in fact, it is even not yet clear 
whether \binf\ is non-empty. These issues will be dealt with in subsections
\ref{s2} to \ref{sinw} below, where we will in particular see that the set 
\binf\ is countably infinite. However, what we already can see is 
that the set \binf\ is linearly independent. To prove this, consider any set 
of finitely many distinct elements $\psi_i$, $i=1,2,...\,,N$ of \binf.
We first show that to any pair $i,j\in\{1,2,...\,,N\}$ there exists a height
$h_{ij}\iNi$ such that\\[.2em]
\mbox{~~~~~}(i)~~$\psi_i(h_{ij})\ne 0$ \,and\, $\psi_j(h_{ij})\ne 0$ \,\ \ 
and\\[.2em]
\mbox{~~~~~}(ii)~$\psi_i(h_{ij})\ne \pm\psi_j(h_{ij})$\,.\\[.2em]
To see this, assume that the statement is wrong, i.e.\ that for each height
$h$ either one of the elements $\psi_i(h)$ and $\psi_j(h)$ of \rh\
vanishes, or one has $\psi_i(h)= \pm\psi_j(h)$. Now because of $\psi_i\ne0$
and $\psi_j\ne0$ there exists heights $h_i$ and $h_j$ with $\psi_i(h_i)\ne0$
and $\psi_j(h_j)\ne0$. This implies that also $\psi_i(\tilde h_{ij})\ne 0$
and $\psi_j(\tilde h_{ij})\ne 0$ for $\tilde h_{ij}:=h_ih_j$.
By our assumption it then follows that $\psi_j(\tilde h_{ij})= \pm
\psi_i(\tilde h_{ij})$, which in turn implies that 
$\psi_j(h_i)= \pm \psi_i(h_i)\ne0$.
Now this conclusion actually extends to arbitrary heights $h$. 
Namely, from the previous result we know that for any $h$ the elements
$\psi_i(h\tilde h_{ij})$ and $\psi_j(h\tilde h_{ij})$ must both be non-zero. 
By our assumption this implies that $\psi_j(h \tilde h_{ij})=\pm\psi_i(h \tilde
h_{ij})$. Projecting this equation down to the height $h$, it follows that
$\psi_j(h)=\pm\psi_i(h)$. Since $h$ was arbitrary,
it follows that in fact $\psi_j= \pm\psi_i$,
and hence (as $-\psi_i$ is not in \binf) that $\psi_j=\psi_i$. This is in
contradiction to the requirement that all $\psi_i$ should be distinct.
Thus our assumption must be wrong, which proves that (i) and (ii) are fulfilled.

Applying now the properties (i) and (ii) for any pair $i,j\in\{1,2,...\,,N\}$
with $i\ne j$, it follows that at the height $h:=\prod_{i,j;i<j}h_{ij}$
\futnot{It is already sufficient to take $h$ as the s.c.m.\ of the $h_{ij}$.}
we have\\[.2em]
\mbox{~~~~~}(i)~~$\psi_i(h)\ne 0$ \,for all\, $i=1,2,...\,,N$ \,\ \ and\\[.2em]
\mbox{~~~~~}(ii)~$\psi_i(h)\ne \pm\psi_j(h)$ \,for all $i,j\in\{1,2,...\,,N\}$,
$i\ne j$\,.\\[.2em]
Thus all the elements $\psi_i(h)$ of \rh\ are distinct and, up to sign, 
elements of the distinguished basis \bah. This implies in particular that 
the only solution of the equation $\sum_{i=1}^N\xi_i\psi_i(h)=0$ is
$\xi_i=0$ for $i=1,2,...\,,N$, which in turn shows that also the equation 
$\sum_{i=1}^N\xi_i\psi_i=0$ has only this solution.
Thus, as claimed, the $\psi_i$ are linearly independent elements of \rinf.

\subsection{\binf\ generates all of \rinf}\label{s2}

Next we claim that the set \binf\ spans \rinf\ in the sense
that the closure (in the \topo) of the linear span of \binf\ in \rinf, i.e.\ 
of the set
  \be  \sbinf \equiv {\rm span}_\zet^{}(\binf)  \ee
of finite \Zet-linear combinations of elements of \binf, is already all of
\rinf.

To prove this, assume that the statement is wrong, or in other words, that the 
set
  \be  \mw := \rinf\,\setminus\,\sbinfc \labl{mw}
is non-empty. By definition, the set $\mw$ is open, and hence because of the
\prepro\ it contains a subset $\mm\subseteq\mw$ of the form $\mm=\efm h(M)$
for some $h\iNi$ and some $M\subseteq\rh$. Further, 
  as an immediate consequence of the construction that we will present
in the subsections \ref{siw} and \ref{sinw}, for each $a\iN\ph$ there exists
an element (in fact, infinitely many elements) $\psi_a\iN\binf$ such that
$\ef h(\psi_a)=\vphih_a$ (namely, we need to prescribe the value of
$\psi_a(p)$ only for the finitely many prime factors $p$ of $h$).
Now choose some $y\iN M$, decompose it \wrtt basis \bah\ of \rh, i.e.
$y=\sumpH a n_a\,\vphih_a$ with $n_a\iN\zet$, and define
$\mx:=\sumpH a n_a\,\psi_a$. Then, on one hand, by construction we have
$\ef h(\mx)=y$, i.e.\ $\mx\iN\mm$, and hence $\mx\in\mw$, while on the other 
hand $\mx$ is a finite linear combination of elements of \binf\
(since $y$ is a finite linear combination of elements of \bah), and hence
$\mx\in\sbinf\subseteq\sbinfc$. By the definition \erf{mw} of $\mw$, this
is a contradiction, and hence our assumption must be wrong.

Together with the result of the previous subsection we thus see that 
\binf\ is a (topological) basis of \rinf.

\subsection{Distinguished sequences of integral weights}\label{siw}

We will now construct all elements of the projective limit \rinf\ 
which belong to the subset \binf\ as introduced in subsection \ref{s1}.
These are obtained as generalized sequences $\psi$ satisfying both \erf j
and the defining relation \erf{bd} of \binf.
More specifically, we construct sequences ${(\ah)}_{h\ini}$ of labels
$\ah\iN\ph$ and associated signs $\ets(\ah)$ such that 
all those $\psi$ which are of the form
  \be  \psi(h)=\ets(\ah)\,\vphih_{\ah} \labl Q
belong to the subset $\binf\subset\rinf$. When applied to \erf Q,
the requirement \erf j amounts to
  \be  \fflh(\vphilh_\alh) = \ets(\alh)\ets(\ah)\cdot\vphih_\ah  \,, \labl k
which in view of the definition \erf' of \fflh\ is equivalent to
  \be  \alh=w(\ah) \qquad{\rm for\ some}\ w\iN\wh    \labl m
and
  \be  \ets(\alh)\ets(\ah)=\epsl(\alh) \,. \labl M

To start the construction of the elements of \binf, we first concentrate our
attention to integral weights of the height $h$ theory which are not
necessarily integrable and which are considered as defined only modulo 
$h\lv$; we denote these weights by $\bh$. Suppose then that we prescribe for
each prime $p$ such a weight $\bp$ and that these weights satisfy in
addition the restriction that for any two primes $p,\,p'$ they differ by
an element of the coroot lattice,
  \be  \bp-\bpp \in \lv \,.  \labl A
We claim that there then exists a sequence ${(\bh)}_{h\ini}$ which for 
prime heights takes the prescribed values $\bp$ and for which the relation
  \be  \bhp = \bh \mod h\lv  \labl F
holds for all $h,\,h'$ with $h\l h'$.

To prove this assertion, we display such a sequence explicitly. To this end, let
  \be  h=:\prodjh\pjnj  \labl G  
denote the decomposition of $h$ into prime factors, and define
  \be  \hi:=\frac{h}\pini   \labl H
and
  \be  \hii:= (\hi)^{-1}_{} \mod \pini  \,. \labl J
Then we set
  \be  \bh := \bpe + \sumieh \hi\,\hii\,(\bpi-\bpe) \,. \labl{bh}
Recall that $\bh$ is defined only modulo $h\lv$.
In \erf{bh} $p_1$ is any of the prime divisors of $h$; it has been singled out 
only in order to make the formula for $\bh$ to look as simple as possible, and 
in fact $\bh$ does not depend (modulo $h\lv$) on the choice of $p_1$. 
To see this, let $\bhz$ denote the number obtained analogously as in
\erf{bh}, but with $p_1$ replaced by some other prime factor $p_2$ of $h$. Then
  \be  \begin{array}{l}  \bh-\bhz=\bpe-\bpz + \sumiezh \hi\,\hii\,(\bpz-\bpe)
  \\[1.5mm] \hsp{19.1} + \hz\,\hzz\,(\bpz-\bpe) - \he\,\hee\,(\bpe-\bpz)
  \,.  \end{array}\ee
Using the fact that $\hi$ is divisible by $\pjnj$ for all primes $p_j$ 
dividing $h$ except for $j=i$, and that $\hi\hii=1\mod\pini$,
it is easily checked that the right hand side of this expression
vanishes modulo $\pjnj\lv$ for all $p_j$ dividing $h$, and hence, using \erf A,
also vanishes modulo $h\lv$.

To establish the coherence property \erf F, we now consider 
two heights $h,\,h'$ such that $h|h'$. Then we set
  \be  h'=:\prodjhp\pjnjp   \ee
and $\hip:=h'/\pinip$, and without loss of generality we can assume that $p_1$ 
divides $h$ as well as $h'$. By the definition \erf{bh} we then have
  \be  \bhp-\bh= \sumieh \Llb \hip\,\hiip - \hi\,\hii \Lrb \,(\bpi-\bpe)
  +\!\!\! \sumihh \hip\,\hiip \,(\bpi-\bpe) \,;  \ee
again it is straightforward to verify that this
vanishes modulo $\pjnj\lv$ for all $p_j$ dividing $h$. This shows that the 
property \erf F is satisfied for the sequence defined by \erf{bh} as claimed.  

Next we note that we did not require that the prescribed values $\bp$
lie on the Weyl orbit of an integrable weight at height $p$, but rather
they may also lie on the boundary of some Weyl chamber of \wp. However,
if $\bp$ does belong to the Weyl orbit of an integrable weight, then also
each weight $\bh$ with $p|h$ is on the Weyl orbit of an integrable weight at 
height $h$. Namely, because of the property \erf F we have in particular
  \be  \bh = \bp + p^n\,\beta  \ee
for some $\beta\iN\lv$. Hence, assuming that $\bh$ is left invariant by
some $w\iN\wh$, i.e.\ that $\bh=w(\bh)\equiv\barw(\bh)+h\gamma$ for some
element $\gamma$ of the coroot lattice, it follows that
  \be  \begin{array}{l}  \barw(\bp)= \barw(\bh-p^n\beta) =
  \barw(\bh)-p^n\,\barw(\beta) \\[1.5mm] \hsp{9.9} 
  = \bh - h\gamma -p^n\,\barw(\beta)
  = \bp +p^n\,(\beta-\barw(\beta))- h\gamma \,.  \end{array}\ee
Since by assumption the only element of \wp\ which leaves the weight
$\bp$ invariant is the identity, it follows that $\barw=\id$ and
$\gamma=0$, implying that also the only element of \wh\ that leaves 
$\bh$ invariant is the identity, which is equivalent to the claimed property.

Our next task is to investigate to what extent the sequence ${(\bh)}_{h\ini}$ 
is characterized by the prescribed values $\bp$ at prime heights and
by the requirement \erf F. To this end let ${(\Bh)}_{h\ini}$ be another
such sequence, i.e.\ a sequence such that $\Bp=\bp$ for all primes
$p$ and $\Bhp-\Bh\in h\lv$ for $h|h'$.
First we observe that for $h$ and $h'$ coprime, the properties
  \be  \Bhh = \Bh \mod h\lv \qquad{\rm and}\qquad \Bhh = \Bhp \mod h'\lv  \ee
fix $\Bhh$ already uniquely (modulo $hh'\lv$), so that the whole freedom
is parametrized by the freedom in the choice of $\Bh$ at heights which
are a prime power. Concerning the latter freedom, we claim that for any prime 
$p$ there is a sequence of elements $\bepj$ of the coroot lattice $\lv$ which
are defined modulo $p\lv$ such that the most general choice of $\Bpn$ reads
  \be  \Bpn = \bpn + \sumne j \bepj\,p^j  \labl{bb}
with $\bpn$ defined according to \erf{bh}, i.e.\ simply $\bpn=\bp$. This 
statement is proven by induction. For $n=1$ it is trivially fulfilled. 
Further, assuming that \erf{bb} is satisfied for some $n\ge1$ and setting 
$\gamma:=\Bpne-\bpne$ (defined modulo $p^{n+1}\lv$), one has
  \be  \Bpne-\Bpn= \gamma-\sumne j \bepj\,p^j \,.  \ee
By the required properties of the sequence $(\Bh)$, the left hand side
of this formula must vanish modulo $p^n\lv$, and hence we have
  \be  \gamma = \bepn\,p^n + \sumne j \bepj\,p^j = \sumn j \bepj\,p^j  \ee
for some $\bepn\iN\lv$.
This shows that $\Bpne$ is again of the form described by \erf{bb};  
furthermore, as $\gamma$ is defined modulo $p^{n+1}\lv$, $\bepn$ is defined
modulo $p\lv$ as required, and hence the proof of the formula \erf{bb} is
completed.

With these results we are now in a position to give a rather explicit 
description of the allowed sequences $\Bh$. Namely, we can parametrize the 
general form of $\Bh$ in terms of the freedom in $\Bpn$ according to
  \be  \begin{array}{l}  \hsp{-3}\Bh= \Bpene + \sumieh\hi\,\hii\,(\Bpini-\Bpene)
  \\[1.5mm] \hsp{.7}
  = \bpe + \sumnee j \bepej\,p_1^j + \sumieh \hi\,\hii\,(\bpi-\bpe) 
    + \sumieh \hi\,\hii\,\llb \sumnie j \bepij\,p_i^j - \sumnee j \bepej\,p_1^j 
  \lrb
  \\[1.5mm] \hsp{.7}
  = \bh + \sumieh \hi\,\hii\,\llb \sumnie j \bepij\,p_i^j - \sumnee j 
    \bepej\,p_1^j \lrb + \sumnee j \bepej\,p_1^j \,.
  \end{array} \labl X
Further, any such sequence fulfills the consistency requirement that 
$\Bhp-\Bh\in h\lv$ for heights $h,\,h'$ with $h|h'$. Namely, in this case the
formula \erf X yields
  \be  \begin{array}{l}  \hsp{-3.8} \Bhp-\Bh = \sumieh \Llb 
  \hip\,\hiip\, \llb \sumnipe j \bepij\,p_i^j - \sumnepe j \bepej\,p_1^j \lrb
  - \hi\,\hii\, \llb \sumnie j \bepij\,p_i^j - \sumnee j \bepej\,p_1^j \lrb \Lrb
  \\[1.5mm] \hsp{16.4}
  + \sumihh
  \hip\,\hiip\, \llb \sumnipe j \bepij\,p_i^j - \sumnepe j \bepej\,p_1^j \lrb
  + \sumneee j \bepej\,p_1^j  \ \mod h\lv \,.
  \end{array} \labl Y
Once more one can easily check that this expression vanishes modulo $\pjnj\lv$
for all primes $p_j$ that divide $h$, and hence vanishes modulo $h\lv$. Thus
the consistency requirement indeed is satisfied.

\subsection{Distinguished sequences of integrable weights and the basis \binf}
\label{sinw}

What we have achieved so far is a characterization of all sequences
${(\bh)}_{h\ini}$ of integral weights defined modulo $h\lv$ that satisfy \erf F.
We now use this result to construct sequences ${(\ah)}_{h\ini}$ of 
highest weights which satisfy the requirement \erf m and of which infinitely
many are {\em integrable\/} weights, with all non-integrable weights being
equal to zero.
We start by prescribing integrable weights $\ap\iN\pp$ for all primes $p$
with $p\ge\gv$, and set $\bp=\ap$ for $p\ge\gv$, while for all primes
$p<\gv$ we choose arbitrary weights $\bp$ (which are necessarily 
non-integrable). Next we employ the previous results to find the sequences
${(\bh)}_{h\ini}$. Finally we define $\ah$ for any arbitrary height $h$ as
follows. If $\bh$ lies on the boundary of a Weyl chamber \wrt \wh, then
we set $\ah=0$. Otherwise there are a unique element $\barw_h\in\barW$ and a
unique
\futnote{To be precise, because the weights $\ah$ are defined only modulo $h\lv
$, $\gamma_h$ is only unique once a definite representative of the equivalence
class of weights that is described by $\ah$ is chosen.}
element $\gamma_h\iN\lv$ such that $\barw_h(\bh)+h\gamma_h$ is integrable, and
in this case we set
  \be  \ah:= \barw_h(\bh)+h\gamma_h \,. \ee
By construction, the weights $\ah$ have the following properties.
If $\ahp=0$ for some height $h'$, then $\bhp$ is on the boundary of a Weyl
chamber \wrt \whp; for any $h$ dividing $h'$, it then follows from
$\bhp-\bh\iN h\lv$ that $\bh$ is on the boundary of a Weyl chamber \wrt \wh,
and hence we also have $\ah=0$. On the other hand, if
$\bhp$ is equivalent \wrt \whp\ to an integrable weight $\ahp\iN\php$, then it 
is a fortiori equivalent to $\ahp$ \wrtt larger group \wh, and then the property
$\bhp-\bh\iN h\lv$ implies that also $\bh$ is equivalent \wrt \wh\ to $\ahp$,
and hence that the associated weight $\ah$ is integrable at height $h$ and is
equivalent \wrt \wh\ to $\ahp$, too. Thus $\ah$ and $\ahp$ are on the same orbit
\wrt \wh\ whenever $h$ divides $h'$, and hence \erf m holds as promised.

Note that by construction for all \gb\ except $\gb=A_1$ the sequences so
obtained contain some zero weights. However, any sequence which contains at 
least one non-zero weight contains in fact infinitely many non-zero (and
hence integrable) weights.

The final step is now to define 
  \be  \psi(h):=\ets(\ah)\,\vphih_{\ah}  \labl x
as in \erf Q, where $\ah$ is as constructed above, and where
  \be  \ets(\ah):= \left\{ \begin{array}{ll} \sign(\barw_h) & {\rm for}\ 
  \ah\iN\wh\,, \\[1mm] 0 & {\rm for}\ \ah=0 \,. 
  \end{array} \right. \labl{ets}
To show that $\psi$ is an element of the projective limit, it only remains
to check the property \erf M of the prefactor $\ets(\ah)$. For $\ah=0$ \erf M 
just reads $0=0$ and is trivially satisfied. For $\ah\iN\ph$, the previous 
results show that $\alh=w_{\el h}^{}\circ w_0\circ w_h^{-1}(\ah)$, where $w_0$ 
is the Weyl translation relating $\bh$ and $\blh$, so that
  \be  \epsl(\alh) = \sign(w_{\el h}^{}\circ w_0\circ w_h^{})
  = \sign(\barw_{\el h}^{})\cdot\sign(\barw_h^{}) \,. \ee
In view of the definition \erf{ets} of $\ets(\ah)$, this is precisely the
required relation \erf M. We conclude that the basis \binf\ of \rinf\ precisely 
consists of the elements \erf x.  In particular, \binf\ is countably infinite.

\Sect{The fusion ring of \mbox{$\bar{\mathfrak g}$}}{gb}

As already pointed out in the introduction, it is expected that in the
limit of infinite level of \wzwts\ somehow the simple \lie\ \gb\ which is the
\hsa\ of \g\ and its \rep\ theory should play a r\^ole. More specifically,
one might think that the representation ring \rb\ of \gb\ emerges. As
we will demonstrate below, indeed this ring shows up, but it is only a proper
subring of the projective limit \rinf\ we constructed, and almost all 
elements of \rinf\ are {\em not\/} contained in the ring \rb.

Let us describe \rb\ and its connection with the category \fusg\ in some 
detail. \rb\ is defined as the ring
over \Zet\ of all isomorphism classes of \findim\ \gb-\rep s, with the ring
product the ordinary tensor product of \gb-\rep s (or, equivalently, the
pointwise product of the characters of these \rep s).
This ring \rb\ is a fusion ring with an infinite basis. 
The elements $\vphib_a$ of a distinguished basis of \rb\ are labelled by
the (shifted) highest weights of irreducible \findim\ \gb-\rep s, i.e.\
by elements of the set
  \be  \pb := \{ a\in\lw \mid 0<a^i\ {\rm for}\, \ i=1,2,...\,,r \}  \,.
  \labl{pb}
Now for any $h\iN\I$ let us define the map $\fb h\!:\ \rb\to\rh\,$ as follows.
If $a\iN\pb$ lies on the boundary of some Weyl chamber \wrt \wh, we set
$\fb h(\vphib_a):=0$; otherwise there exist a unique $a'\iN\ph$ 
and a unique $w\iN\wh$ such that $w(a)=a'$, and in this case we set
  \be  \fb h(\vphib_a):= \eps(a)\cdot \vphih_{a'}   \labl`
with $\eps(a)=\sign(w)$. As in the case of the maps $\fflh$ \erf', we
will consider \erf` as covering all cases, i.e.\ set $\eps(a)=0$ if
$a$ lies on the boundary of a Weyl chamber at height $h$.

To analyze the relation between the ring \rb\ and the category \fusg,
we first recall the expressions
  \be  \nbabc =  \sum_{\barw\in\barW} \sign(\barw)\,\mult b(\barw(c)-a)
  \labl{nb}
for the \lrc s (or tensor product coefficients) of \gb\ \cite{raca2,spei} and
  \be  \nhabc = \sum_{w\in\wH} \sign(w)\,\mult b(w(c)-a)
  \labl{nh}
for the \frc s, i.e.\ the structure constants of the WZW fusion ring \rh\
\cite{KAc3,walt3,fugp2,fuva2}. Here $\mult a(b)$ denotes the multiplicity of 
the (shifted) weight $b$ in the \gb-\rep\ with (shifted) highest weight $a$.
It will be convenient to extend the validity of \erf{nb} by adopting it
as a definition of \nbabc\ for arbitrary (i.e., not necessarily lying in 
\pb) integral weights $a$ and $c$, and also extend it to arbitrary integral 
weights $b$ that do
not lie on the boundary of any Weyl chamber \wrt $\barW$ by setting
  \be  \mult b(c):=\sign(\barw_b^{})\,\mult{\barw_b^{}(b)}(c)  \,, \ee
with $\barw_b$ the unique element of $\barW$ such that $\barw_b(b)\iN\pb$.

The multiplicities $\mult a(b)$ are invariant under the Weyl group $\barW$,
i.e.\ $\mult a(\barw(b))=\mult a(b)$ for all $\barw\iN\barW$. 
As a consequence, the numbers \nbabc\ and \nhabc\ are related by
\futnote{In the formulation of \cite{KAc3,walt3,fugp2,fuva2} the factor of
$|\barW|^{-1}$ is absent because there \nbabc\ is taken to be non-zero only
if $a,b\iN\pb$.}
  \be  \nhabc = \frac1{|\barW|}\, \sum_{w\in\wH} \sign(w)\,\nb ab{w(c)} \,.
  \labl B
The invariance of $\mult a(b)$ under $\barW$ also implies
that for arbitrary integral weights $a,\,b$ and $c$ the symmetry property
  \be  \nbabc = \nb bac  \labl{sy}
follows from the analogous property of the \lrc s with $a,b,c\in\pb$, and that
  \be  \begin{array}{l}  \nb{\barw_1(a)}b{\,\ \barw_2(c)} 
  = \dsum_{\barw\in\barW} \sign(\barw)\,\mult b(\barw\,\barw_2(c)-\barw_1(a))  
  \\{}\\[-3mm] \hsp{16.2}
  = \dsum_{\barw\in\barW} \sign(\barw)\,\mult b(\barw_1^{-1}\barw\:\barw_2(c)-a)
  = \sign(\barw_1\barw_2)\cdot\nbabc \,. \end{array}  \ee
When combined with the symmetry property \erf{sy}, the latter formula yields
  \be  \nb{\barw_1(a)}{\barw_2(b)}{\ \ \ \ \ \ \barw_3(c)} 
  = \sign(\barw_1\barw_2\barw_3)\cdot\nbabc \,.  \ee

To obtain information about the effect of affine Weyl transformations on 
the labels of \nbabc, we consider an alternating sum over the Weyl group \wh. 
We have
  \be  \bearl  \dsum_{w_2\in\wH}\!\!\sign(w_2)\,\nb{w_1(a)}b{\ \,w_2(c)} 
  = \!\!\dsum_{\scriptstyle\barw,\barw_2\in\barW \atop\scriptstyle\beta_2\in\lV}
  \!\!\sign(\barw)\sign(\barw_2)
  \,\mult b(\barw\,\barw_2(c)+h\barw(\beta_2)-\barw_1(a)-h\beta_1)
  \\{}\\[-3mm] \hsp{43.7}
  = \dsum_{\barw,\barw_2\in\barW} \dsum_{\beta\in\lV} \sign(\barw\,\barw_2)
  \,\mult b(\barw_1^{-1}\barw\,\barw_2(c)+h\barw_1^{-1}\barw(\beta)-a)
  \\{}\\[-3mm] \hsp{43.7}
  = \sign(w_1)\cdot\dsum_{w_2\in\wH}\sign(w_2)\,\nb ab{w_2(c)} \,.
  \end{array}\ee
Here $\beta:=\beta_2-\barw^{-1}(\beta_1)$.
Together with the symmetry property \erf{sy} it then follows that
  \be  \sum_{w_3\in\wH}\sign(w_3)\,\nb{w_1(a)}{w_2(b)}{\ \ \ \ \ w_3(c)} 
  = \sign(w_1)\,\sign(w_2)\cdot\sum_{w_3\in\wH}\sign(w_3)\,\nb ab{w_3(c)} 
  \labl C
for all $w_1,w_2\iN\wh$. We can rewrite this as
  \be  \sum_{w\in\wH}\epsl(A)\epsl(B)\,\sign(w)\,\nb{w_A(A)}{w_B(B)}
  {\ \ \ \ \ \ \ \ \ \ w(c)}
  = \dsum_{w\in\wH} \sign(w)\,\nb AB{w(c)} \,  \labl{515}
which by interpreting $A$ and $B$ as elements of \pb\ rather than \plh\ 
yields, after summation over $c\iN\ph$,
  \be  \fb h(\vphib_A)\rdot\fb h(\vphib_B)=\fb h(\vphib_A\rdot \vphib_B) \,, \ee
and hence shows that the maps \fb h defined by \erf` are ring homomorphisms.

Now for all $A\iN\plh$ we have $\fflh(\phi_A)=\epsl(A)(\phi_{w_A(A)})
=:\epsl(A)\cdot \vphi_a$. Then owing to \erf B we obtain, after dividing 
\erf{515} by $|\barW|$, the relation
  \be  \nh{\!\FFlh(A)}{\FFlh(B)}{\ \ \ \ \ \ \ \ \ \ \ \ \ c}
  = \epsl(A)\,\epsl(B)\,\nhabc
  = \dsum_{C:\ \phi_C\in\FFlh^{-1}(\varphi_c)} \epsl(C)\;\nlh ABC   \labl:
(on the \lhs, we use the short hand notation $\fflh(A)\iN\ph$ to indicate
the label that corresponds to the element $\epsl(A)\fflh(\phi_A)$ of \rh).
Summation over $c\iN\ph$
then yields $\fflh(\phi_A)\rdot\fflh(\phi_B)=\fflh(\phi_A\rdot \phi_B)$,
so that \erf: is just the homomorphism property of the maps \ff ij which were
defined by \erf' in terms of the \frc s. (Thereby we have also obtained 
an alternative proof of the homomorphism property of those maps.)

To investigate further the relation between $\rb$ and the projective limit
\rinf, we introduce the linear mappings
  \be  \begin{array}{llll}  \jb h:& \rh\to\rb\,,& \vphih_a\mapsto\vphib_a \,,
  \\[1.6mm]  \jj h{h'}:& \rh\to\rhp\,,& \vphih_a\mapsto\vphihp_a &
  \end{array} \labl{jj}
which map each basis element $\vphih_a$ of \rh\ to that basis element of \rb\ 
and \rhp\ ($h'\ge h$), respectively, which is labelled by the same weight
$a\in\ph\subseteq\php\subset\pb$.
For $h\l h'\l h''$, these maps satisfy
  \be  \jb{h'}\circ\jj h{h'} =\jb h \,, \quad
  \jj{h'}{h''}\circ\jj h{h'} =\jj h{h''}\, \ee
as well as
  \be  \fb{h}\circ\jb h =\id_h \,, \quad
  \ff{h'}{h}\circ\jj h{h'} =\id_{h} \ee
and
  \be  \fb{h}\circ\jb{h'} =\ff{h'}h \,, \quad
  \ff{h''}{h}\circ\jj{h'}{h''} =\ff{h'}h \,. \ee

We say that a generalized sequence $\psi$ in the projective limit \rinf\ is 
{\em ultimately constant\/} iff there exists a $\ho\iNi$ such that 
  \be  \psi(h) = \jj\ho h\circ\psi(\ho) \labl{const} 
(and hence for basis elements in particular $\ah= a_\ho$) for all heights 
$h\ge\ho$. Now assume that $\psi_1$ and $\psi_2$ are elements of \rinf\ 
which are ultimately constant, with associated heights $\hoe$ and $\hoz$, \resp.
Then in particular for all heights $h$ larger than $\ho:=2\,{\rm max}
(\hoe,\hoz)$ the fusion product $\psi_1(h)\rdot\psi_2(h)$ in \rh\
is isomorphic to the product $\psib_1\rdot\psib_2$ in \rb, where $\psib_1
:=\jb\ho\circ\psi_1(\ho)$, and analogously for $\psib_2$. This implies that
  \be  (\jj\ho h\circ\psi_1(\ho)) \rdot ((\jj\ho h\circ\psi_2(\ho))
  = \jj\ho h\circ (\psi_1(\ho) \rdot\psi_2(\ho))   \labl{labl}
even though $\jj\ho h$ is not a ring homomorphism, and hence
$(\psi_1\rdot\psi_2)(h)\equiv\psi_1(h)\rdot\psi_2(h)=\jj\ho h\circ
(\psi_1\rdot\psi_2)(\ho)$ for all $h\ge\ho$. Thus the product
$\psi_1\rdot\psi_2$ is again ultimately constant. Also, the property of being
ultimately constant is preserved upon taking (finite) linear transformations and
conjugates. The set of ultimately constant elements therefore constitutes
a subring of \rinf.
\futnot{positivity of the $\nabc$ in the subring is guaranteed once the over-all
sign for the basis elements is chosen properly.}

The following consideration shows that this subring is isomorphic to the
fusion ring \rb. First, any ultimately constant element is a linear
combination of ultimately constant elements $\psia$ for which $\psia(\ho)$
is an element of the canonical basis of \rHo, $\psia(\ho)=\vphiho_a$ for
some $a\in\pho\subset\pb$.
But there is a unique element $\psi$ of \rinf\ with the latter property,
because at all heights $h$ smaller than $\ho$ the value $\psi(h)$ is
already fixed by imposing the requirement \erf k. Thus there is a bijective 
linear map between the subring of ultimately constant elements and the
fusion ring \rb, defined by $\vphib_a\mapsto\psia$ for $a\iN\pb$.
Moreover, the same argument which led to \erf{labl} shows that this
map is in fact an isomorphism of fusion rings. As this map is provided
in a canonical manner, we can actually identify the two rings.

A generic element of \rinf\ is {\em not\/} ultimately constant, so that
the subring of ultimately constant elements is a proper subring of \rinf.
Thus what we have achieved is to identify the fusion ring \rb\ as a
proper sub-fusion ring of the projective limit ring \rinf.

To conclude this section, let us remark that of course we could have enlarged 
by hand the category \fusg\ to a larger category \fusgb\ by just including 
one additional object into the category, namely the ring \rb, together with
the morphisms \fb h. This essentially amounts to cutting the category 
of rings in  such a way that one is able to
identify the ring \rb\ as the projective limit of this category \fusgb.
We do not regard this as a viable alternative to our construction, though,
since when doing so one performs manipulations which are suggested merely
by one's prejudice on what the limit should look like. (Also, phenomena like 
level-rank dualities in fusion rings require to consider various rings for 
different \alg s \g\ on the same footing; the category \fusgb\
cannot accommodate such phenomena.)
In contrast, our construction of the limit employs only the description in
terms of coherent sequences, which is a natural procedure for any small
category, and does not presuppose any desired features of the limit.

\Sect{\Rep\ theory of $\rinf$}{rep}

A basic tool in the study of fusion rings is their \rep\ theory. Of particular
importance are the \irrep s, which lead in particular to the notion of 
(generalized) quantum dimensions. In this section
we show that an analogous \rep\ theory exists for the projective limit as
well. In our considerations the \topo\ will again play an essential r\^ole.

\subsection{\Onedim\ \rep s}

Let us consider for any two $h,h'\iNi$ with $h'=\el h$ the 
injection of the label set \ph\ (defined as in \erf{ph}) into the label set 
\php\ that is defined by multiplying the weights $a$ by a factor of $\el$: 
  \be  a \;\mapsto\; \el a  \ee
for all $a\iN\ph$. (This induces an injection $\vphih_a\mapsto\vphilh_{\el a}$ 
of the distinguished basis $\bah$ of $\rh$ into the distinguished basis $\balh$ 
of $\rlh$. However, when this map is extended linearly to all of \rh, it does
{\em not\/} provide a homomorphism of fusion rings.)
We can use these injections to perform an {\em inductive\/} limit 
of the set $(\ph)_{h\in I}$ of label sets, where the set $I$ \erf i is again 
considered as directed via the partial ordering \erf l. We denote this 
inductive limit by $\indu$.
An element $\alpha$ of $\indu$ can be characterized by an integrable 
weight $\halpha\iN\ph$ at some suitable height $h$; at any multiple $\el h$
of this height, the same element $\alpha$ of $\indu$ is then represented
by the weight $\lhalpha=\el\,\halpha$. In particular, quite unlike as in 
the case of the projective limit, each element of the inductive limit $\indu$ 
is already determined by its representative at a single height. Also note that 
an element $\alpha\iN\indu$ is {\em not\/} defined at all heights $h$;
in particular, for any $h\iNi$ the set of those $\alpha\iN\indu$ 
which have a representative at height $h$ is in one-to-one correspondence with 
the elements of \ph, and hence is in particular finite.
We will use the notation $\ainph$ to indicate that $\alpha\iN\indu$ has a 
representative $\halpha\iN\ph$ at height $h$.

We claim that any element of $\indu$ gives rise to a one-dimensional \rep\ of
the projective limit \rinf\ of the fusion rings. To see this, we choose 
for a given $\alpha\iN\indu$ a suitable height $h\iNi$ such that $\ainph$.
To any coherent sequence $(\psi(l))_{l\in I}$ in the projective 
limit $\rinf$ we then associate the number
  \be \cald_\alpha(\psi) := 
  \frac{\sh_{\psi(h),\halpha}}{\sh_{\hrho,\halpha}} \,,  \labl{cald}
i.e.\ the $\halpha$th quantum dimension of the element $\psi(h)$ of the ring 
\rh. Here we use the short-hand notation
  \be  \sh_{\psi(h),b}:=\sumph a \zeta_a\, \sh_{a,b} \qquad {\rm for}\;\
  \psi(h)=\sumph a \zeta_a\,\vphih_a  \ee
for linear combinations of \smat\ elements.
Using the identities \erf D and \erf{sS} as well as $\lhalpha=\el\halpha$
and the defining properties of $\psi$, we have
  \be  \frac{\slh_{\psi(\el h),\lhalpha}}{\slh_{\lhrho,\lhalpha}}
  = \frac{(\slh\,\DD)_{\psi(\el h),\halpha}}{(\slh\,\DD)_{\lhrho,\halpha}}
  = \frac{(\PP\,\sh)_{\psi(\el h),\halpha}}{(\PP\,\sh)_{\lhrho,\halpha}}
  = \frac{\sh_{\psi(h),\halpha}}{\sh_{\hrho,\halpha}}  \,; \labl{s/s}
this shows that the formula \erf{cald} yields a well-defined map from \rinf\ 
to \Complex\,, i.e.\ it does not depend on the particular choice of $h$.
Using the knowledge about the \rep\ theory of the rings \rh, it 
then follows immediately that
  \be  \cald_\alpha(\psi)\,\cald_\alpha(\psip) = \sumph{a,b} \zeta_a^{}
  \zeta_b'\,\nhabc\, \frac{\sh_{c,\halpha}}{\sh_{\hrho,\halpha}}
  = \cald_\alpha(\psi\rdot\psip) \,.  \ee
Thus the prescription \erf{cald} indeed provides us with a \onedim\ \rep\ of 
$\rinf$.

Let us now associate to any element $\psi$ of $\rinf$ the infinite sequence of
quantum dimensions \erf{cald}, labelled by $\indu$; this way we obtain a map
  \be \cald:\quad \psi \,\mapsto\, (\cald_\alpha(\psi))_{\alpha\in\indu}
  \labl{ca}
from the ring \rinf\ to the algebra 
  \be  \diago := \{ (\xi_\alpha)_{\alpha\in\indu} \Mid \xi_\alpha\iN\complex\,
  \} \ee
of all countably infinite sequences of complex numbers.
Since we are now dealing with complex numbers rather than only integers, it
is natural to consider instead of the fusion ring \rinf\ the corresponding
algebra over \Complex\,, to which we refer as the {\em fusion algebra\/} \ainf.
(For simplicity we regard \ainf\ as an algebra over \Complex\,.
In principle it would be sufficient to consider it over a certain 
subfield of \Complex\ generated by appropriate roots of unity.) It is then
evident that the map $\cald\!:\,\ainf\to\diago$ defined by \erf{ca} is an 
algebra homomorphism. (We continue to use the symbol $\cald$. More generally,
below we will always assume that the various maps to be used, such as
the projection \erf3, are continued \Complex\,-linearly from the fusion rings
\rh\ to the associated fusion algebras \Ah, and use the same symbols for
these extended maps as for the original ones.)

\subsection{An isomorphism between \ainf\ and $\diago$}

In this subsection we show that the map $\cald$ introduced above
even constitutes an 
{\em iso\/}morphism between the complex algebras \ainf\ and $\diago$: 
  \be  \cald:\quad  \ainf \;\stackrel{\cong}{\longrightarrow}\; \diago \,. 
  \labl{dax}

Injectivity of $\cald$ is easy to check. Suppose we have $\cald(\psi)=0$. Fix 
any $h\iN I$; then all quantum dimensions of the element $\psi(h)$ of \rh\
vanish. From the properties of the fusion ring \rh\ it then follows 
immediately that $\psi(h)=0$. This is true for all $h\iN I$, and hence we have
$\psi=0$. This proves injectivity.

To show also surjectivity requires more work. We first need to introduce the 
elements
  \be  \idem_a \equiv \idemh_a := \sh_{\rho,a} \sumph b \sh^*_{a,b}\, \vphih_b 
  \;\in \Ah \labl{ea}
of the fusion algebras at height $h$. These elements are idempotents, i.e.\ obey
  \be  \idem_a\rdot \idem_b = \delta_{a,b}\, \idem_{a} \,. \labl{eee}
Owing to the unitarity of the modular 
transformation matrix $S$, the idempotents $\{\idemh_a\Mid a\iN\ph\}$ form a 
basis of the fusion \alg\ \Ah, and they constitute a partition of the unit 
element, in the sense that
  \be  \sumph a \idemh_a = \vphih_\rho  \,.  \labl{pou}
Also, for any element $\psi\iN\ainf$ with 
$\psi(h)=\idem_a$ and any $\alpha\iN\indu$ with $\ainph$ we have
  \be  \cald_\alpha(\psi) = \delta_{a,\halpha} \,.  \labl{pea}

We now study how the idempotents $\idem_{\halpha}$ behave
under the projection \erf3. First, when $\alpha\iN\indu$ has a
representative $\halpha$ at height $h$, then for every positive
integer $\el$ we have, using the first of the identities \erf{sS},
  \be \begin{array}{ll}  \fflh(\idem_{\lhalpha}) \!\!
  &= {\slh_{\rho,\lhalpha}} \sumplh A \!\!\slh^*_{\lhalpha,A}\fflh(\phI_A)  
   = {\slh_{\rho,\el\halpha}} \sumplh A\, \sumph b \! \slh^*_{\lhalpha,A}
      \P Ab\, \vphi_b  \\{}\\[-.3em]
  &= \el^{-r/2}\cdot {\sh_{\rho,\halpha}} \sumph b\!(\slh^* \PP)_{\lhalpha,b}
      \, \vphi_b
   = {\sh_{\rho,\halpha}} \sumph{b,c} \! D_{\el\halpha,c}\sh^*_{c,b}\, \vphi_b
  \\{}\\[-.7em]
  &= \sh_{\rho,\halpha} \sumph b \sh^*_{\halpha,b}\, \vphi_b
   = \idem_{\halpha} \, .  \end{array}\labl{ee}
On the other hand, when $\alpha$ has a representative at height $h$, but
not at height $h'$, we can compute as follows. Since $hh'$ is a multiple of
$h$, $\alpha$ has a representative $\hhpalpha$ at height $hh'$. Thus
we can repeat the previous calculation to deduce that
   \be  \bearll \ff{hh'}{h'}(\idem_{\hhpalpha}) \!\!&
   = \shhp_{\rho,\hhpalpha} \cdot h^{r/2} \sumphp b (\DD\,\shp)_{\hhpalpha,b} \,
   \vphihp_b \\{}\\[-.7em]  & = (h/h')^{r/2}\cdot \sh_{\rho,\halpha} 
   \sumphp{b,c} \delta_{h'\halpha,hc} \shp_{c,b}\, \vphihp_b  \,.  \eear \ee
Now in the sum over $c$ on the \rhs\ one has a contribution only if 
$c=h'\halpha/h=\hhpalpha/h$ is an element of the label set \php\ at height $h'$.
But in this case we would conclude that $\alpha$ has in fact a representative
at height $h'$, namely $\hpalpha=c$, which contradicts our assumption.
Therefore we conclude that in the case under consideration we have
$\ff{hh'}{h'}(\idem_{\hhpalpha})=0.$ Together with the result \erf{ee}
it follows that by setting
  \be  \eal(h):= \left\{ \bearll  0 & \mbox{if $\alpha\iN\indu$ has no
  representative at height $h$}\,, \\[.1em] \idem_{\halpha} & {\rm else}\,,
  \eear \right.  \labl{eal}
we obtain an element $\eal$ of the projective limit \ainf.
Moreover, according to the relation \erf{pea} the map \erf{dax} acts on
$\eal\iN\ainf$ as
  \be  \lLb \cald(\eal) \lRb_\beta = \delta_{\alpha,\beta}    \labl{peb}
for all $\alpha,\beta\iN\indu$, and the $\eal$ provide a partition of the unit 
element, analogously as in \erf{pou}, 
  \be  \sum_{\alpha\in\indu} \eal = \psio  \,;  \ee
here the sum is to be understood as a limit of finite sums in the \topo.

Now for each $h\iNi$ let us define the map \,$g_h\!:\, \diago\to\Ah$ by
$(\xi_\alpha)_{\alpha\in\indu}\mapsto \sum_{\scriptstyle\beta\in\indu 
\atop \scriptstyle\Hbeta\iin\pH} \xi_\beta\, \idem_{\hbeta}$.
Since $\Lhbeta\iiN\plh$ \,if\, $\Hbeta\iiN\ph$, we then have 
  \be  \fflh \circ g_{\el h}^{}((\xi_\alpha)) = 
  \sum_{\scriptstyle\beta\in\indu \atop \scriptstyle\Lhbeta\iin\plH}\xi_\beta\, 
  \idem_{\beta}(h) = 
  \sum_{\scriptstyle\beta\in\indu \atop \scriptstyle\Lhbeta\iin\pH}\xi_\beta\, 
  \idem_{\hbeta} = g_h((\xi_\alpha))  \ee
for all positive integers $\el$. Analogously we can define a map
  \be  g:\quad \diago \to \ainf\,,\qquad (\xi_\alpha)_{\alpha\in\indu}^{}\mapsto
  \!\!\sum^{}_{\beta\in\indu} \xi_\beta\,\idem_\beta  \ee
with similar properties. As a consequence of the relation \erf{peb} one finds 
that this map satisfies
  \be  \cald \circ g = \id_{\diago}^{} \,. \ee
This implies that the injective map $\cald$ is also surjective (and that 
$g$ is injective). Thus we have proven the isomorphism \erf{dax}.

\subsection{Semi-simplicity}

It is known \cite{kawA} that the fusion \alg s \Ah\ at finite heights $h$
are \ssi\ associative \alg s. In this subsection we show that in a suitable
topological sense the same statement holds for the projective limit \ainf, too.

We first combine the identity \erf{eee} and the definition \erf{eal}
of the element $\eal$ of \ainf\ with the fact that the idempotents 
$\idem_a$ form a basis of \Ah. This way we learn that for all $\psi\iN
\ainf$ and all heights $h\iNi$ the fusion product $(\eal\rdot\psi)(h)=
\eal(h)\rdot\psi(h)$ is proportional to $\eal(h)$. 
Thus for each $\alpha\iN\indu$ the span
  \be  \Id_\alpha :=  \eals   \labl{Id}
of $\{\eal\}$ is a \onedim\ twosided ideal of the projective limit, i.e.\ we 
have $\ainf\,\Id_\alpha=\Id_\alpha\ainf\subseteq\Id_\alpha$.
We claim that when we endow the \alg\ with the \topo, then in fact \ainf\ 
is the closure of the direct sum of the ideals \erf{Id} in this topology:
  \be  \ainf = \overline{\bigoplus_{\alpha\in\indu} \Id_\alpha} \,. \labl{clos}
(In particular, the idempotents $\eal$ form a topological basis of \ainf.)

To prove this, we first recall from subsection \ref{topo} that in the \topo\ 
each open set in \ainf\ is a union of elements of the set 
$\Om=\{ \efm h(M) \Mid h\iN\I,\; M\;{\rm open\;in}\;\Ah\}$
of all pre-images of all open sets in any of the fusion \alg s \Ah.
Here we assume that we have already chosen a topology on each of the fusion
\alg s \Ah. (Actually the choice of this topology on \Ah\ will not be important;
for definiteness, we may take the discrete one, as in the case of fusion rings,
or also the metric topology of \Ah\ as a \findim\ complex vector space.)
Consider now an arbitrary element $\xi\iN\ainf$, which because of the 
isomorphism $\ainf\cong\diago$ we can write as $\xi=(\xi_\alpha)_{\alpha\in
\indu}$. Adopting some definite numbering $\indu=\{\alpham\Mid m\iN\natnum\}$
of the countable set $\indu$, for $n\iN\natnum$ we define 
  \be  \xih_n := \sum_{m\le n} \xi_\alpham \ealm \ \in\; \bigoplus_{\alpha\in
  \indu} \Id_\alpha \,.  \ee
To prove our assertion, we must then show that
for every $h\iNi$ and every open set $M\subseteq\Ah$ which satisfy
$f_h(\xi)\iN M$ we have $\xih_n\iN f_h^{-1}(M)$, i.e.\
  \be  f_h(\xih_n) \in M \,,  \ee
for all but finitely many $n$. Now by direct calculation we obtain
  \be  f_h(\xih_n) = \sum_{\scs m\le n \atop\ \scs\alpham\iin\pH} 
  \!\! \xi_\alpham \idemh_\halpham \,; \ee
this is a finite sum, and for sufficiently large $n$ it becomes independent
of $n$ because only finitely many $\alpha\iN\indu$ have a representative
in \ph. In fact, for sufficiently large $n$ we simply have
  \be  f_h(\xih_n) = \sum_{\alpha\iin\pH} \xi_\alpha\, \idemh_\halpha
  \equiv f_h(\xi) \,.  \ee
Since $f_h(\xi)\iN M$, this immediately shows that indeed $f_h(\xih_n) \iN M$
for almost all $n$, and hence the proof is completed. (Note that the fact that
$f_h(\xih_n)$ ultimately becomes equal to $f_h(\xi)$ holds for any chosen
topology of \Ah, and hence the conclusion is indeed independent of that
topology.)

\subsection{Simple and \ssi\ modules}\label{Sss}

The \rep\ theory of \ainf\ can now be developed by following the
same steps as in the \rep\ theory of \ssi\ algebras.
However, when considering modules $V$ over \ainf, 
it is natural to restrict one's attention from the outset to 
{\em continuous\/} modules, i.e.\ to modules which
are topological vector spaces and on which the \rep\ of \ainf\ is continuous 
(in particular, every element of \ainf\ is represented by a continuous map).
We will do so, and suppress the qualification `continuous' from now on.

The \onedim\ ideals $\Id_\alpha$ are simple modules over \ainf\ under the (left
or right) regular \rep. Our first result is now that these \onedim\ modules
already provide us
with all simple modules, i.e.\ that every simple \ainf-module $L$ satisfies
  \be  L \,\cong\,\Id_\alpha \ee
for some $\alpha\iN\indu$. 

To show this, we first observe that if $L\not\cong\Id_\alpha$, then
$\Id_\alpha L = 0$. Namely, since $\Id_\alpha$ is an ideal of \ainf, we have 
$\ainf\,\Id_\alpha L\subseteq\Id_\alpha L$; thus $\Id_\alpha L$ is a submodule 
of $L$, which by the simplicity of $L$ implies that either
$\Id_\alpha L = L$ or $\Id_\alpha L=0$. In the former case,
$\Id_\alpha L=L$, we can find a vector $y\iN L$ such that the space 
$\Id_\alpha y$ is not zero-dimensional. Indeed, because of
$\ainf\,\Id_\alpha\,y\subseteq \Id_\alpha\,y\subseteq L$
this space is a submodule of $L$, and hence by the simplicity of $L$ it must
be equal to $L$. It follows that the map from $\Id_\alpha$ to $L$ defined by
$\lambda \mapsto \lambda y$ is surjective. Since $L$ is simple, by Schur's lemma
this implies that it is even an isomorphism. This shows that $L\cong\Id_\alpha$
when $\Id_\alpha L=L$, and hence $\Id_\alpha L=0$ when $L\not\cong\Id_\alpha$.\\
Suppose now that $L$ is a non-zero simple module and is not isomorphic to any 
$\Id_\alpha$. Then $\bigoplus_\alpha\Id_\alpha L = 0$; since $L$ is a 
continuous module, we can take the closure of this relation, so as to find that 
$\ainf\,L =0$. But we have $L\subseteq\ainf\,L$, and hence
this would imply that $L=0$, which
is a contradiction. Hence we learn that indeed, up to isomorphism,
the ideals $\Id_\alpha$ of \ainf\ exhaust all the simple modules over \ainf.

Next we consider modules $V$ over \ainf\ which can be obtained 
from families of simple modules. Similarly as in \cite[\S XVII.2]{LAng}
one can show that the following conditions are equivalent:
\begin{quote}
$(i)$~~~~$V$ is the closure of the sum of a family of simple submodules.\\[.3em]
$(ii)$~~~$V$ is the closure of the direct sum of a family of simple submodules.
\\[.3em]
$(iii)$\hsp{2.85}Every closed submodule $W$ of $V$ is a direct summand, i.e.\ 
there exists\\\hsp{10.48}a closed submodule $W'$ such that $V = W \oplus W'$.
\end{quote}
Any (continuous) module fulfilling these equivalent conditions 
will be referred to as a {\em\ssi\/} module. 

The equivalence of $(i)$\,\hy\,$(iii)$ is proven as follows.
First, if $V=\overline{\sum_{i\in J}L_i}$ is the closure of a (not necessarily 
direct) sum of simple submodules $L_i$, denote by $J'$ a maximal subset of 
$J$ such that $V':=\sum_{j\in J'}L_j$ is a direct sum. Since the
intersection of $V'$
with any of the simple modules $L_i$ is a submodule of $L_i$, the maximality
of $J'$ implies that $i\iN J'$ and hence in fact $J'=J$. 
Thus $(i)$ implies $(ii)$. \\
Second, if $W$ is a submodule of $V$, let $J''$ be the maximal subset of 
$J$ such that the sum $W+\sum_{j\in J''}L_j$ is direct. Then the same
arguments as before show that $V = \overline{W \oplus \bigoplus_{j\in J''}L_j}$.
If, furthermore, $W$ is closed, then it follows that $V = \overline{W}\oplus 
\overline{\bigoplus_{j\in J''}L_j} = W\oplus\overline{\bigoplus_{j\in J''}L_j}$.
This shows that$(ii)$ implies $(iii)$.\,%
\futnote{It is indeed necessary to require $W$ to be closed. Consider e.g.\ 
the case $V=\ainf$ and $W=\bigoplus_\alpha\Id_\alpha$. 
The submodule $W$ is neither closed nor does it have a complement.}

\noindent Third, assume that $V$ is a non-zero module which satisfies $(iii)$,
and let $v$ be a non-zero vector in $V$. 
The kernel of the homomorphism $\ainf\to\ainf\, v$ is a closed ideal of \ainf,
which in turn is contained in a maximal closed ideal ${\cal J}\!\subset\!\ainf$ 
that is strictly contained in \ainf. One then has $V={\cal J}v \oplus W$
and $\ainf\,v={\cal J}v \oplus (W\cap\ainf\, v)$ with some submodule
$W\subset V$. Now $W\cap\ainf\, v$ is simple because ${\cal J}v$ is maximal
in $\ainf\, v$; thus $V$ contains a simple submodule. Next let $V' \ne 0 $ be 
the submodule of $V$ that is the closure of the sum of all simple
submodules of $V$. If $V'$ were not all of $V$, then one would have
$V=V'\oplus V''$ with $V''\ne0$; but by the same reasoning as before,
$V''$ then would contain a simple submodule, in contradiction to the definition
of $V'$. Thus $V'=V$, so we see that $(iii)$ implies $(i)$.

\subsection{Arbitrary modules}

With the characterization of \ssi\ modules above, we are now in a position
to study arbitrary modules of \ainf, in an analogous manner as
in \cite[\S XVII.4]{LAng}. Let us first assume that $W$ is a closed submodule 
of a \ssi\ module $V$, and denote by $W'$ the closure of the 
direct sum of all simple submodules of $W$. Then there is a submodule
$V'$ of $V$ such that $V= W'\oplus V'$. Every $w\iN W$ can be uniquely
written as $w=w'+v'$ with $w'\iN W'$ and $v'\iN V'$. Because of
$v'=w-w'\iN W$ we thus have $W=W'\oplus(W\cap V')$. The module
$W\cap V'$ is a closed submodule of $W$. If it were non-zero, it would
therefore (by the same reasoning as in the proof of `$(iii)\to(i)$' in
subsection \ref{Sss}) contain a simple submodule, in contradiction with 
the definition of $W$. Thus we learn that $W=W'$, or in other words:
\begin{quote}
Every closed submodule of a \ssi\ \ainf-module is \ssi.
\end{quote}

\noindent
Next we consider again a closed submodule $W$ of a \ssi\ module $V$, and
investigate the quotient module $V/W$.
There is a closed submodule $W'$ such that $V$ is the direct sum
$V=W\oplus W'$. Now the projection $V \to V/W$ induces a continuous
isomorphism from $W'$ to $V/W$. Furthermore, according to the result 
just obtained, $W'$ is \ssi. Thus we have shown:
\begin{quote}
Every quotient module of a \ssi\ \ainf-module with respect to \\
a closed submodule is \ssi.
\end{quote}

\noindent Now 
any arbitrary \ainf-module can be regarded as a quotient module of a suitable 
free module modulo a closed submodule. Moreover, every free \ainf-module is the 
closure of a direct sum of countably many copies of \ainf\ and hence is a \ssi\ 
module. The two previous results therefore imply:
\begin{quote}
Every \ainf-module is \ssi. 
\end{quote}

Finally we consider again an arbitrary \ainf-module $V$. We denote by $V_\alpha$
the closure of the direct sum of all those submodules of $V$ which are
isomorphic to the simple \ainf-module $\Id_\alpha$. Since each
simple module over \ainf\ is isomorphic to some
$\Id_\alpha$, any simple submodule of $V$ is contained in $V_\beta$ for some 
$\beta\iN\indu$. Now every \ainf-module is \ssi\ and hence the closure 
of the direct sum of its simple submodules. Thus we learn that
  \be  V = \overline{ \bigoplus_{\alpha\in\indu} V_\alpha } \, . \ee
Moreover, we have $\idem_\beta V_\alpha=\delta_{\alpha,\beta}V_\alpha$ for
all $\alpha,\beta\iN\indu$, and hence
  \be  V_\alpha = \idem_\alpha V = \Id_\alpha V \, . \ee
As a consequence, we see that:
\begin{quote}
Every \ainf-module $V$ can be written as
  \be  V = \overline{\bigoplus_{\alpha\in\indu} \Id_\alpha V} = 
  \overline{\bigoplus_{\alpha\in\indu} \idem_\alpha V} \,, \ee
and for each $\alpha\iN\indu$ the submodule $\Id_\alpha V$ is the closure of 
the direct sum of all submodules of $V$ that are isomorphic to $\Id_\alpha$. 
\end{quote}

\noindent
We can conclude that the structure of any arbitrary (continuous)
module over \ainf\ is known explicitly, i.e.\ we have developed the full 
(topological) \rep\ theory of \ainf.

\subsection{Diagonalization}

{}From the definition \erf{eal} of $\eal\iN\ainf$ and the basic property 
\erf{eee} of the idempotents $\idem_a\iN\Ah$ it follows that the elements 
$\eal$ of \ainf\ are again idempotents:
  \be  \eal \rdot \idem_\beta = \delta_{\alpha,\beta}\,\eal  \ee
for all $\alpha,\beta\iN\indu$.
In other words, by the basis transformation from the distinguished basis
\binf\ of the fusion \alg\ \ainf\ to the basis of idempotents one 
diagonalizes the fusion rules of \ainf,
precisely as in the case of the \alg s \Ah\ at finite level.

Indeed, by combining the definitions \erf{ea} and \erf{eal} we can describe
the transformation from \binf\ to the basis of idempotents $\eal$ explicitly. 
Namely, for any $\psi=(\psi(h))_{h\in\II}\iN\binf$ we have 
  \be  \psi = \sum_{\alpha\in\indu}\! \U_{\psi,\alpha} \, \eal \,, \ee
with
  \be  \U_{\psi,\alpha}:= \frac{\sh_{\psi(h),\halpha}}{\sh_{\rho,\halpha}}
  \,, \ee
where $h\iNi$ is a height at which $\alpha$ has a representative. Note that
owing to the relation \erf{s/s} the quotient $\U_{\psi,\alpha}$
does not depend on the particular choice of $h$.
(This just rephrases the fact that the map $\cal D$ is an isomorphism.)

For any $\psi,\psip\iN\binf$ we thus have 
  \be \psi\rdot\psip= \sum_{\alpha\in\indu} 
  \frac{\sh_{\psi(h),\halpha}} {\sh_{\rho,\halpha}}
  \frac{\sh_{\psip(h),\halpha}}{\sh_{\rho,\halpha}} \, \idem_\alpha
  = \sum_{\chi\in \Binf} \LLb \!\!\sum_{\alpha\in\indu} 
  \U_{\psi,\alpha} \U_{\psip,\alpha} \, \UM_{\alpha,\chi}\, \LRb \, \chi \,, \ee
where $\UM$ is the matrix for the inverse basis transformation,
  \be  \eal = \sum_{\psi\in \Binf} \UM_{\alpha,\psi}\, \psi \,. \ee
In other words, the \frc s of the projective limit \ainf\ can be written as
  \be  _{}^{\sss(\infty)}\!\!\n\psi{\psip}{\psipp} = \sum_{\alpha\in\indu} 
  \U_{\psi,\alpha}\, \U_{\psip,\alpha}\, \UM_{\alpha,\psipp} \, . \ee
This is nothing but the analogue of the Verlinde formula \cite{verl2} that 
is valid for the \frc s of the fusion \alg s \Ah.

Note that already at finite height $h$ the two indices which label the rows
and columns, \resp, of the matrix $\sh$ which diagonalizes the fusion rules
are a priori of a rather different nature. Namely, one of them labels the
elements of the distinguished basis \bah, while the other labels the
inequivalent \onedim\ \irrep s of \Ah. It is a quite non-trivial 
\futnot{and not well understood}
property of the fusion \alg s which arise in rational \cft\ 
(and is a prerequisite for the modularity of those fusion \alg s)
that nevertheless the diagonalizing matrix can be chosen such that it is 
symmetric, so that in particular the two kinds of labels can be treated on an
equal footing \cite{jf24}.
Our results clearly display that this nice feature of the finite height
fusion \alg s \Ah\ is not shared by their non-rational limit \ainf; in the
case of \ainf, there seems to be no possibility to identify the two sets
\binf\ and $\indu$ which label the elements of the distinguished basis and
the \onedim\ \irrep s, \resp, with each other.

On the other hand, our results show that the projective limit \rinf\ that
we constructed in this paper still possesses all those structural properties
of a modular fusion ring which can reasonably be expected to survive in 
the limit of infinite level.

 \def\wb{\,\linebreak[0]} \def\wB {$\,$\wb}
 \newcommand\Bi[1]    {\bibitem{#1}}
 \newcommand\Erra[3]  {\,[{\em ibid.}\ {#1} ({#2}) {#3}, {\em Erratum}]}
 \newcommand\BOOK[4]  {{\em #1\/} ({#2}, {#3} {#4})}
 \newcommand\J[5]     {\ {\sl #5}, {#1} {#2} ({#3}) {#4} }
 \newcommand\Prep[2]  {{\sl #2}, preprint {#1}}
 \def\jf    {J.\ Fuchs}
 \def\foph  {Fortschr.\wb Phys.}
 \def\hepa  {Helv.\wb Phys.\wB Acta}
 \newcommand\npbF[5] {{\sl #5}, \nupb\ {#1} [FS{#2}] ({#3}) {#4}}
 \def\npbp  {Nucl.\wb Phys.\ B (Proc.\wb Suppl.)}
 \def\nuci  {Nuovo\wB Cim.}
 \def\nupb  {Nucl.\wb Phys.\ B}
 \def\phlb  {Phys.\wb Lett.\ B}
 \def\jomp  {J.\wb Math.\wb Phys.}
 \def\comp  {Com\-mun.\wb Math.\wb Phys.}
 \def\A       {Algebra}
 \def\AP     {{Academic Press}}
 \def\AW     {{Addi\-son\hy Wes\-ley}}
 \def\alg     {algebra}
 \def\Be     {{Berlin}}
 \def\BIR    {{Birk\-h\"au\-ser}}
 \def\Ca     {{Cambridge}}
 \def\CUP    {{Cambridge University Press}}
 \def\furu    {fusion rule}
 \def\GB     {{Gordon and Breach}}
 \newcommand{\inBO}[7]  {in:\ {\em #1}, {#2}\ ({#3}, {#4} {#5}), p.\ {#6}}
 \def\Infdim  {Infinite-dimensional}
 \def\KLU    {{Kluwer Academic Publishers}}
 \def\NY     {{New York}}
 \def\Q       {Quantum\ }
 \def\qg      {quantum group}
 \def\SV     {{Sprin\-ger Verlag}}
 \def\syms    {sym\-me\-tries}
 \def\trfo    {transformation}
 \def\wzw     {WZW\ }

\vskip5em
\small
\noindent{\bf Acknowledgement.} \ We are grateful to I.\ Kausz and B.\ Pareigis
for helpful comments.

\version\versionno
\ifnum\draftcontrol=0 \def\END{\end{document}} \else \def\END{\newpage} \fi
\END